\documentclass[%
 reprint,
superscriptaddress,
 amsmath,amssymb,
 aps,
]{revtex4-2}

\usepackage{graphicx}
\usepackage{dcolumn}
\usepackage{bm}

\usepackage{xcolor}

\newcommand{\ba}{\begin{eqnarray}}
\newcommand{\ea}{\end{eqnarray}}
\renewcommand{\v}[1]{{\bf #1}}

\begin{document}

\preprint{APS/123-QED}

\title{Non-Abelian topological defects and strain mapping in 2D moir\'e materials}

\author{Rebecca Engelke}
\affiliation{Department of Physics, Harvard University, Cambridge, Massachusetts 02138, USA}
\author{Hyobin Yoo}%
\affiliation{Department of Physics, Sogang University, Seoul 04107, Republic of Korea}
\affiliation{Institute of Emergent Materials, Sogang University, Seoul 04107, Republic of Korea}
\author{Stephen Carr}
\affiliation{Department of Physics, Brown University, Providence, Rhode Island, 02912, USA}
\author{Kevin Xu}
\affiliation{Department of Physics, Harvard University, Cambridge, Massachusetts 02138, USA}
\author{Paul Cazeaux}
\affiliation{Department of Mathematics, Virginia Tech, Blacksburg, Virginia 24061, USA}
\author{Richard Allen}
\affiliation{Department of Physics, Harvard University, Cambridge, Massachusetts 02138, USA}
\author{Andres \surname{Mier Valdivia}}
\affiliation{John A. Paulson School of Engineering and Applied Sciences, Harvard University, Cambridge, Massachusetts 02138, USA}
\author{Mitchell Luskin}
\affiliation{School of Mathematics, University of Minnesota, Minneapolis, Minnesota 55455, USA}
\author{Efthimios Kaxiras}
\affiliation{Department of Physics, Harvard University, Cambridge, Massachusetts 02138, USA}
\affiliation{John A. Paulson School of Engineering and Applied Sciences, Harvard University, Cambridge, Massachusetts 02138, USA}
\author{Minhyong Kim}
\affiliation{International Centre for Mathematical Sciences, Edinburgh, UK}
\author{Jung Hoon Han}
\affiliation{Department of Physics, Sungkyunkwan University, Suwon 16419, Korea}
\author{Philip Kim}
\affiliation{Department of Physics, Harvard University, Cambridge, Massachusetts 02138, USA}%
\affiliation{John A. Paulson School of Engineering and Applied Sciences, Harvard University, Cambridge, Massachusetts 02138, USA}

\date{\today}

\begin{abstract}
We present a general method to analyze the topological nature of the domain boundary connectivity that appeared in relaxed moir\'e superlattice patterns at the interface of 2-dimensional (2D) van der Waals (vdW) materials. At large enough moir\'e lengths, all moir\'e systems relax into commensurated 2D domains separated by networks of dislocation lines. The nodes of the 2D dislocation line network can be considered as vortex-like topological defects. We find that a simple analogy to common topological systems with an $S^1$ order parameter, such as a superconductor or planar ferromagnet, cannot correctly capture the topological nature of these defects. For example, in twisted bilayer graphene, the order parameter space for the relaxed moir\'e system is homotopy equivalent to a punctured torus. Here, the nodes of the 2D dislocation network can be characterized as elements of the fundamental group of the punctured torus, the free group on two generators, endowing these network nodes with non-Abelian properties. Extending this analysis to consider moir\'e patterns generated from any relative strain, we find that antivortices occur in the presence of anisotropic heterostrain, such as shear or anisotropic expansion, while arrays of vortices appear under twist or isotropic expansion between vdW materials. Experimentally, utilizing the dark field imaging capability of transmission electron microscopy (TEM), we demonstrate the existence of vortex and antivortex pair formation in a moir\'e system, caused by competition between different types of heterostrains in the vdW interfaces. We also present a methodology for mapping the underlying heterostrain of a moir\'e structure from experimental TEM data, which provides a quantitative relation between the various components of heterostrain and vortex-antivortex density in moir\'e systems.

\end{abstract}

\maketitle



\section{\label{sec:level1}Introduction}

Moir\'e patterns are quasi-periodic in-plane projections of two similar stacked 2-dimensional (2D) periodic lattices. Atomic scale moir\'e superlattices can be formed by stacking atomically thin van der Waals (vdW) materials; one such example is twisted bilayer graphene. 
Moir\'e patterns formed by incommensurately stacking 2D materials have been used to manipulate a system's electronic structure, from Hofstadter’s butterfly~\cite{HofPonomarenko,HofDean,HofHunt} to the valley Hall effect~\cite{ValleyGorb,ValleyEndo} to magic angle strongly correlated physics~\cite{CaoMott,CaoMagic}. As the number and type of layers in experimentally relevant systems proliferates, including twisted double bilayer~\cite{Xiaomeng-TDBG}, 
twisted trilayer~\cite{Ke,ZeyuTrilayer,TrilayerPark} and twisted quadrilayer graphene~\cite{Quadrilayer}, as well as hexagonal boron nitride~\cite{FE2} and transition metal dichacolgenides (TMDs)~\cite{FE3}, it is important to be able to predict the structure in 
vdW stacked combinations of atomic layers. 

Increasing attention has been paid to the effects of strain disorder on the structure and properties of such systems~\cite{StrainReview}. The effect and extent of twist angle disorder in magic angle graphene is an active area of research~\cite{StrainmapSQUID,Strain2}. Strained moir\'e patterns in excitonic systems have been proposed as a way to create 1D quantum wires~\cite{Excitons}. In this paper, we present a generalizable topological interpretation of the structure of moir\'e interfaces that allows for the characterization of arbitrary strain and the proposition of new types of moir\'e patterns. 

A topological description of the moir\'e structure is appealing in part because some of the major features of the structure seem to be fixed once certain boundary conditions, such as total twist angle and strain, are pinned. For large enough moir\'e length, moir\'e systems are known to relax into domains of nearly commensurate alignment, separated by domain walls which can be characterized as dislocation lines~\cite{Yoo}.  The topological connectivity of the network of dislocation lines remains fixed even as the domain lengths become distorted by local strain fields.

The nodes of the network where dislocation lines meet in the relaxed moir\'e system (sometimes known as AA points in graphene or TMD moir\'e) have been referred to as topological defects by Alden et al.~\cite{Alden}, and again by Turkel et al.~\cite{Turkel}, who also emphasize their role in transport as tunable, local concentrations of twist angle. They have also been shown to play the role of defects in electrochemistry~\cite{Kwabena-chemistry}. However, 
we find that the order parameter describing the topology of these defects is rather different from that of the conventional
vortex descriptions in the planar magnet or superfluid, where the order parameter can be described by a complex number of unit length.
The unit length complex order parameter space can be mapped to
a circle ($S^1$), hence the fundamental group is $\pi_1 (S^1 ) = \mathbb{Z}$. The fundamental group is a useful tool in the analysis of topological defects, as defects and their collections can each be mapped to an element of the fundamental group \cite{thouless_book}. For defects with an $S^1$ order parameter, this manifests as an integer winding number \cite{Mermin}. 
Interestingly, we find that the order parameter space describing the relaxation structure of a moir\'e superlattice is not homotopy equivalent to $S^1$ due to the periodic boundary condition of the order parameter imposed by the moir\'e superlattice.
We present a 
mathematical framework, similarly based on the fundamental group, by which the 
2D dislocation network
nodes can be characterized as topological defects. 

Throughout the paper, we refer to the space of order parameters as the configuration space, in keeping with the convention adopted in earlier works~\cite{CarrRelaxation}. For most of this work, we focus on graphene-like moir\'e superlattices, where two different domain types, AB and BA, are separated by dislocation lines meeting at the AA site. The starting point for our new vortex description of the AA nodes is the realization that the configuration space for the graphene-like moir\'e superlattice is the {\it punctured torus}, which is homotopy equivalent to theta space (to be defined later). The fundamental group of the punctured torus is $F_2$, the free group on two generators $a$ and $b$. The generators $a$ and $b$ are noncommuting, so $F_2$ is non-Abelian. The commutator $[a,b] = aba^{-1}b^{-1}$ corresponds to a closed loop around a vortex centered at the AA node of the moir\'e superlattice (its inverse $[b,a]= [a,b]^{-1}$ corresponds to a path around an antivortex). The non-commutativity of the generators is a consequence of the removed point at the AA node, which is physically motivated by the node's high energy barrier in the generalized stacking fault energy~\cite{CarrRelaxation}. 

The domain walls that emerge in the moir\'e superlattice will be color-coded as $R, G, B$, which correspond to the three distinct ways in which the AB stacking order makes a transition to the BA stacking order as the domain wall is crossed. 
The aforementioned generators $a$ and $b$ can be related to the $RGB$ color coding of the domain walls, which are experimentally accessible quantities. Examination of the color distribution of domain walls crossing an arbitrary closed boundary reveals the vortex content enclosed within, according to the non-Abelian vortex theory developed here. The use of the free group language to characterize the topological structure of a moir\'e superlattice is likely to find application in other kinds of superlattices formed from incommensurate stacking of non-hexagonal crystals, with details of the group determined by the material-dependent stacking energy profiles and lattice symmetry.

As mentioned before, the non-Abelian vortices in a moir\'e superlattice have a counterpart in non-Abelian antivortices. We find that the relative strain tensor between two constituent layers determines the vortex/antivortex distribution of the sample.
The mathematical tools to understand configuration space combined with experimental information on the configuration distribution in real space enables us to estimate the strain distribution underlying a moir\'e pattern, allowing for the characterization or engineering of strain distributions in van der Waals heterostructures. 

The remainder of the paper is organized as follows. In Sec. \ref{sec:2} we go over various tools and concepts used in analyzing the strain patterns and the nature of topological defects in the moir\'e superlattice. The notion of theta space as the proper configuration space of the graphene-like moir\'e superlattice is introduced and justified by energetic consideration. Sec. \ref{sec:3} discusses the formal theory of the vortex and antivortex in a moir\'e superlattice using the language of the free group and its generators. The mathematical discussion is followed by Sec. \ref{sec:4} in which the new algebraic formulation of vortices and anti-vortices is employed to identify antivortex formation in a real moir\'e superlattice, by way of a novel method of strain mapping. 
Sec. \ref{sec:5} gives a summary and discussion. Technical details of the theory of vortex algebra are included in Appendix \ref{sec:appendix-1} and Appendix \ref{sec:appendix-2} with the hope that future investigations of moir\'e superlattices, including lattices other than sublattice-symmetric honeycomb lattices, can make use of the type of vorticity formulation presented here. Appendix \ref{sec:appendix-imageprocessing} includes details on the image processing to experimentally measure the lattice displacement.

\section{Order parameter and configuration space for moir\'e superlattice}
\label{sec:2} 

\begin{figure}
\includegraphics[width=\linewidth]{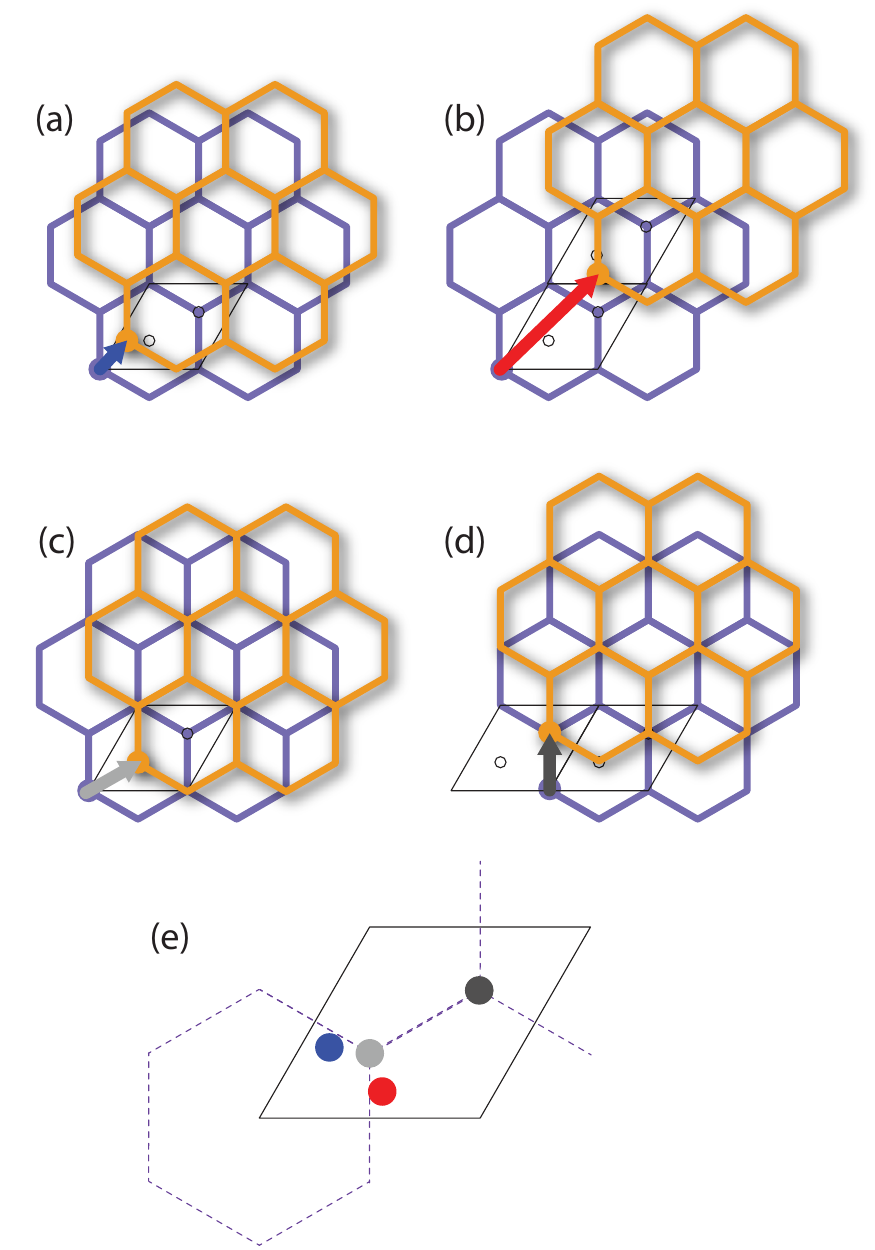}
\caption{\label{fig:shift}(a-d) Definition of shift vector: (a) small shift, (b) shift larger than a unit cell is mapped into the first unit cell, (c), BA shift (d) AB shift.
(e) Shifts from part (a) shown in configuration space. Dotted lines show an alternate Wigner–Seitz-like designation of the configuration space unit cell. }
\end{figure}

The natural choice of order parameter in a bilayer system is the local shift vector, ${\bf{u}}$, determined as the in-plane vector that points from a lattice site in one layer to the equivalent lattice site in the other layer. Fig. \ref{fig:shift} illustrates of how we define the shift vector in a graphene-like honeycomb lattice. Because of the periodicity of the lattice, a shift vector larger than a unit cell, as shown in Fig. \ref{fig:shift}(b), is equivalent to the shorter vector folded into the first unit cell. In other words, the configuration space in which this order parameter exists is a torus~\cite{CarrRelaxation}. 

After labeling the two honeycomb lattice sites as A and B, the standard naming convention for the bilayer honeycomb stackings is obtained by listing the pair of vertically aligned sites.
The condition ${\bf{u}}=0$, when every atom is on top of an equivalent atom in the other layer, is known as AA stacking. Shifting the top layer along one of the three atomic bonds from an A site to a B site of the other layer  (Fig. \ref{fig:shift}(d)), or along the negative of those three vectors (Fig. \ref{fig:shift}(c)) yields a structure where only half the atoms in the top and bottom layers coincide in the 2D projection.  The latter two stacking configurations are often termed AB and BA stacking, respectively. In graphene, the AA stacking is energetically unfavorable, while the AB and BA stackings are symmetry-related lowest-energy layer stacking configurations, called Bernal stacking. Note that for each of the two graphene Bernal stacking configurations, three different shifting directions result in the same stacking configuration, represented by a single point on the toroidal configuration space. 
AB and BA stacking are connected by spatial inversion, leading to ${\bf{u}}_\text{AB} = -{\bf{u}}_\text{BA}$ for the corresponding shift vectors in the configuration space (Fig. \ref{fig:shift}(e)). The high-energy nature of the AA stacking can be reflected by removing ${\bf u} = 0$ from the configuration space altogether, rendering the topology from that of a torus to that of a punctured torus. This will play a crucial role in the theory of vorticity we develop in Sec. \ref{sec:3}. 

Fig. \ref{fig:shift}(e) shows the points in configuration space corresponding to the real-space configurations in (a-d). The high symmetry stackings, BA and AB, are represented by the dark gray and light gray points in Fig. \ref{fig:shift}(e), respectively. As the unit cell can be defined in various ways, it is equally valid to use the parallelogram definition of the unit cell shown in Fig. \ref{fig:shift}, where AA is the corner, or the hexagonal unit cell shown in dotted lines, in which AB and BA are corners of the hexagon. 

\subsection{Experimental measurement of order parameter}
\begin{figure}
\includegraphics[width=\linewidth]{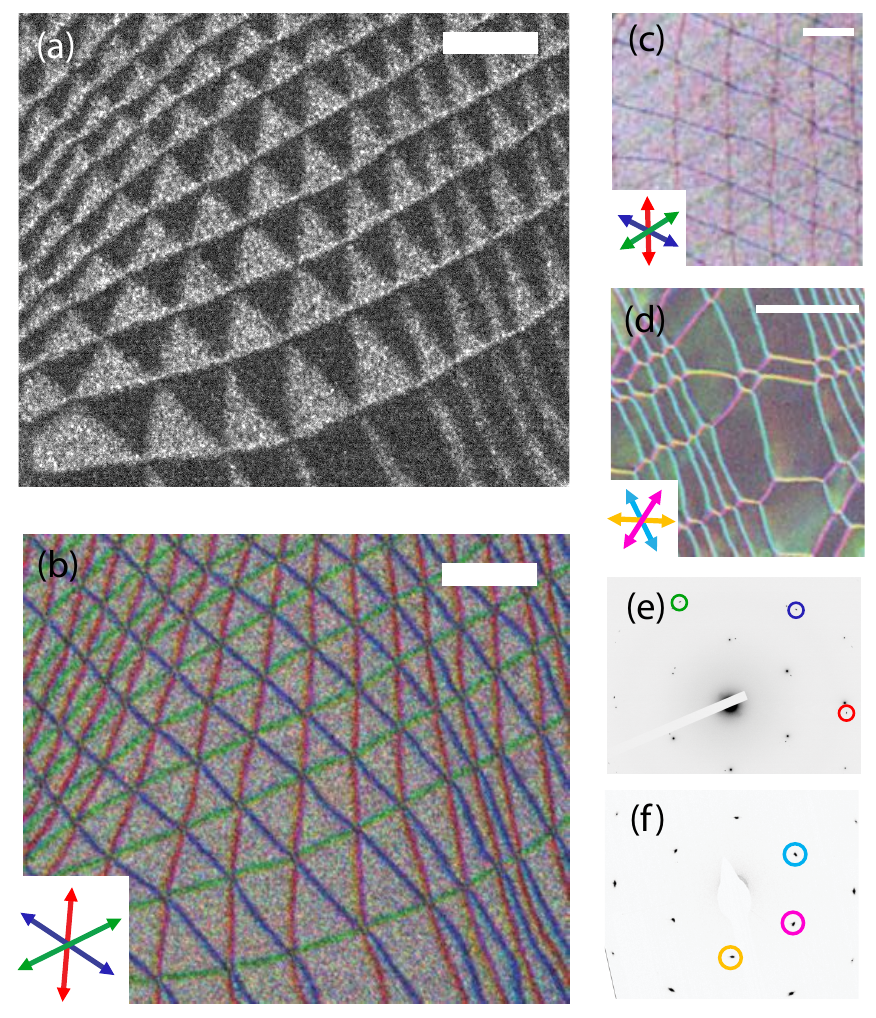} 
\caption{\label{fig:df} Dark field TEM images. (a) $\{1 0 \overline{1} 0\}$ (``first order") dark field of twisted bilayer graphene, showing domain contrast. (b) $\{1 1 \overline{2} 0\}$ (``second order") dark field of the area in (a) showing domain walls meeting at sixfold nodes. Inset: Burgers vectors corresponding to domain wall colors. (c) Second order dark field of MoS$_2$ twisted from 3R-like stacking has a similar network topology to graphene. Inset: Burgers vectors corresponding to domain wall colors. 
(d) $\{1 0 \overline{1} 0\}$ dark field of MoS$_2$ twisted from 2H-like stacking has a threefold network topology. Inset: Burgers vectors corresponding to domain wall colors. (e)  Diffraction pattern of the sample in (a) with colors of second order diffraction spots circled. Sample (c) has a similarly oriented diffraction pattern. (f) Diffraction pattern of the sample in (d) with the colors of first order diffraction spots circled.  Scale bars are 100nm.
}
\end{figure}

Transmission electron microscopy (TEM) provides an experimental route to characterizing local atomic configurations in real space, including detecting the change in $\v u$ across a domain wall, known as the Burgers vector. Changes in the stacking order are distinguishable by a dark field imaging technique that consists of inserting an aperture into the diffraction plane around a single Bragg position and recording the resulting filtered real space image. Depending on the choice of diffraction peak, contrast between domains (see Fig. \ref{fig:df}(a)) or the partial dislocations that form the domain walls (see Fig. \ref{fig:df}(b)) in the bilayer stacking are visible~\cite{Alden}. The Burgers vectors of the dislocations are exactly determinable, as the vector perpendicular to the diffraction peak for which their contrast vanishes in the dark field image~\cite{Alden,LinAC}. The Burgers vector of a dislocation in the bilayer is equal to $\Delta \v u$, the change in shift vector across the boundary. 
 
We note that the dislocations form a network with distinct topology. Twisted bilayer graphene (Fig. \ref{fig:df}(b)) has a structure where six dislocation lines meet at a node.  Despite being a different material with different symmetries, MoS$_2$ slightly twisted from a 3R-like stacking (close to 0$^\circ$ twisting) appears to share the same dislocation network topology as graphene (Fig. \ref{fig:df}(c)). In contrast, MoS$_2$ twisted from a 2H configuration (close to 180$^\circ$ twisting) has a structure where three lines meet at a node rather than six (Fig. \ref{fig:df}(d)). As we will explore later, the topology of the graphene and 3R-like case is defined by a punctured torus, originating from the single energy maximum in the stacking fault energy as a function of configuration (and two degenerate minima). However, for twisted 2H MoS$_2$ there are two energy maxima and one minimum, leading the configuration space to be described as a twice-punctured torus and generating a different topology. While in this paper we focus on the topology generated by the energy profile of bilayer graphene, we are motivated by the realization that a topological description of each system's configuration space should distinguish the different possible connectivities of the network, including those arising from materials of different lattice symmetry, such as a square lattice. 

%

While constructing the tools to analyze the topological defects of the moir\'e superlattice in configuration space, we also attain the ability to analyze the configuration distribution and extract the overall strain profile in the twisted bilayer.
In order to do these analyses, we need to consider several important quantities of the moir\'e lattice: the displacement gradient matrix, the moir\'e vectors, and the Burgers vectors. All of these quantities give important information that allows us to reconstruct ${\bf{u}}$. 

\subsection{Displacement gradient matrix}
A displacement vector field $\bf{u}$ which describes the shift in positions of a lattice $({\bf{x}_t})$ compared to a reference lattice $({\bf{x}_b})$ such that ${\bf{x}_t}={\bf{x}_b}+{\bf{u}(\bf{x}_b)}$, is closely related to strain. Typically, ${\bf{x}_t}$ 
is a 2D vector corresponding to
the positions in the strained lattice, and ${\bf{x}_b}$ 
is a 2D vector corresponding to the positions in the intrinsic lattice, prior to application of forces. In the case of moir\'e materials, presented in Fig. \ref{fig:shift}, the reference lattice (with coordinates ${\bf{x}_b}$) is the bottom layer of the material, and the displaced lattice (${\bf{x}_t}$) is the top layer. Thus the vector field that produces a moir\'e pattern is in a sense analogous to the relative strain (heterostrain) between the two layers.

The linear strain matrix, used in the modeling of strain energies, is defined by taking derivatives of the ${\bf{u}}$ field and symmetrizing:

\begin{equation} \label{strain}
	 \varepsilon = \left(	\begin{array}{cc} 
	\partial_x u_x & (\partial_y u_x+\partial_x u_y)/2  \\
    (\partial_y u_x+\partial_x u_y)/2 & \partial_y u_y    \\ 
	\end{array}
	\right). \end{equation}

Note that this linear strain tensor removes rotation contributions to first order in angle, since rotating the lattice does not cost energy and thus should not be counted towards the total strain energy.

Without symmetrizing, the gradient of the ${\bf{u}}$ field is known as the displacement gradient matrix, $\overline{d}$, obtained from  
	
	\begin{equation} \label{displacementgradient}
	 \overline{d} = \nabla {\bf{u}} = \left( 	\begin{array}{cc}
	\partial_x u_x & \partial_y u_x  \\
	\partial_x u_y & \partial_y u_y  \\
	\end{array}
	 \right). \end{equation}

The vector ${\bf{u}}$ can be recovered by integrating $\overline{d}$ over distance. If we assume that $\bf{u}(0) = 0$ and that the strain is spatially uniform from $\bf{0}$ to $\bf{x}_b$,
we obtain
\begin{equation}
	{\bf{u}(\bf{x}_b)} = \overline{d} {\bf{x}_b}.\label{eq:d} 
\end{equation}
Thus, ${\bf{x}_t} = {\bf{x}_b} + \overline{d} {\bf{x}_b}$ or \begin{equation} {\bf{x}_t} =  (1 + \overline{d}) {\bf{x}_b} \label{eq:transf} \end{equation}

Any arbitrary displacement of layer 2 relative to layer 1, formed by a combination of twist and strain, can be written in terms of a displacement gradient matrix, $\overline{d}$, with four independent components. It is useful to define the four components as follows:
\begin{equation} \label{eq:defd}
    \overline{d}= \left(	\begin{array}{cc}
	\alpha+\beta & \gamma-\theta \\
	\gamma + \theta & \alpha-\beta  \\	
	\end{array}
	\right),
\end{equation}
where $\theta$ is a linear approximate of the twist (measured in radians), $\alpha$ is isotropic strain, and $\beta$ and $\gamma$ are uniaxial and shear strain, respectively~\cite{StrainMatrix}.


\subsection{Moir\'e length}

The moir\'e length is the distance over which the lattice ${\bf{x}_t}$ has been shifted by a unit vector with respect to lattice ${\bf{x}_b}$, creating a local return to the starting configuration. 

Consider the shift vector ${\bf{u}}$, as defined in Fig. \ref{fig:shift}. If we twist our lattices about a point where two atoms are stacked on top of each other, the origin has ${\bf{u}}=0.$ Traveling away from the origin in a direction ${\bf{x}}$, the displacement ${\bf{u}}$ increases until we reach the point where the lattices have diverged by a whole unit cell (${\bf{u}}=\pm{\bf{a}}_1$ or $\pm{\bf{a}}_2$, where ${\bf{a}}_1$ and ${\bf{a}}_2$ are two lattice vectors of the unstrained lattice) and thus are aligned (${\bf{u}}=0)$ again, resulting in moir\'e periodicity. 


A general interface of 2D lattices with heterostrain, ${\bf{x}_t}=(1+\overline{d}){\bf{x}_b}$, 
can be characterized by a pair of moir\'e vectors, ${\bf{m}}_i$ ($i=$1, 2), which  describes the moir\'e periodicity in the 2D space. Considering the corresponding pair of two coincident lattice points in the upper and lower layers, ${\bf{x}_t}_{,i}$ and ${\bf{x}_b}_{,i}$, respectively, the two moir\'e vectors can be expressed 
\begin{equation} \label{eq:Moirelength}
	{\bf{m}}_i={\bf{x}_t}_{,i}={\bf{x}_b}_{,i} +{\bf{s}}_i,  
\end{equation}
where the constant vectors ${\bf{s}}_i$  
can each be $\pm {\bf{a}}_1$ or $\pm {\bf{a}}_2$, depending on the coincident lattice condition for the moir\'e superlattice. 
 
If ${\bf{s}}_1$ and ${\bf{s}}_2$ are collinear (without ${\bf{m}}_1$ and  ${\bf{m}}_2$ being collinear), then the matrix $\overline{d}$ has determinant zero and the moir\'e pattern is 1D rather than 2D. We will ignore this case for now, and assume that ${\bf{a}}_1$ and ${\bf{a}}_2$ are each used once.  



We can relate the lattice constants $\bf{a_i}$, the matrix $\overline{d}$ and the moir\'e vectors, by ${\bf{m}}_i ={\bf{x}_t}_{,i} = (1+\overline{d}){\bf{x}_b}_{,i}.$ Thus if $\overline{d}$ is invertible, ${\bf{x}_b}_{,i}=\overline{d}^{-1}{\bf{s}}_i$ and 

\begin{equation} \label{eq:Moirelength-inv}
	{\bf{m}}_i=(1+\overline{d}^{ -1}){\bf{s}}_i.
\end{equation}

Putting Eq.~\ref{eq:Moirelength-inv} together with the corresponding equation for the other moir\'e vector, ${\bf{m}}_i$, we obtain a matrix equation that can be solved for $\overline{d}$ if we have measured $\overline{m}$ and $\overline{s}$:
\begin{equation} \label{eq:solve-d}
	    \overline{m} =(1+\overline{d}^{-1}) \overline{s},
\end{equation} 
where $\overline{m}$ and $\overline{s}$ are $2\times 2$ matrices formed by horizontally concatenating the ${\bf{m}}_i$ and ${\bf{s}}_i$ column vectors, respectively. Note that if the two ${\bf{m}}_i$'s are not linearly independent, that is again the $|\overline{d}|=0$ case.

\subsection{Burgers vector}
\label{sec:burgers}

The shift vector ${\bf{u}}$ is also related to the Burgers vector of the dislocations that form in the relaxed moir\'e system.
To define the Burgers vector, first one considers a closed path along the lattice points of a perfect crystal, sometimes called a Burgers circuit.
When a dislocation is introduced within this path the circuit becomes broken, and the vector connecting the now separated start and end points of the circuit is called the Burgers vector.
In a moir\'e superlattice, the twisted interface acts as a dislocation, and we can obtain Burgers vectors by considering circuits that traverse the interface. 
The simplest closed circuits in the aligned, untwisted structure take the following form: travel from a lattice point $\bf{x}_0$ to $\bf{x}_1$ along the top layer, then down vertically into the bottom layer, then along $\bf{x}_1$ to $\bf{x}_0$ in the bottom layer, and finally return to $\bf{x}_0$ in the top layer by moving vertically upwards.
To connect the Burgers vector $\bf{b}$ to the shift vector $\bf{u}$, it is easiest to think of obtaining the twisted geometry by keeping the bottom layer fixed and introducing a dislocation via rotation of the top layer.
In this case, the failure of the circuit to close after applying the twist is equal to the relative change in positions of the lattice points that corresponded to $\bf{x}_0$ and $\bf{x}_1$ in the top layer.
However this is simply the change in the local shift between the two points, e.g.

\begin{equation}
\bf{b} = \int_{\bf{x}_0}^{\bf{x}_1} ( d\bf{x} \cdot \bf{\nabla}) \bf{u}(\bf{x}) = \bf{u}(\bf{x}_1) - \bf{u}(\bf{x}_0)
\end{equation}
where the non-integral form is only true if one does not map $\bf{u}$ into the compact unit-cell torus.




While this expression was obtained under the assumption of a uniformly twisted system, the result is quite general.
For the relaxed systems that occur in experimental devices, one finds that the Burgers vector is only non-zero if a ``dislocation line'' (that is, a domain wall) is enclosed in the circuit.
Each dislocation line between a pair of AA nodes can be associated with a specific Burgers vector, as was done experimentally using DF TEM in Fig.~\ref{fig:df}.
The pair of AA nodes also provides a moir\'e vector ${\bf{m}}_i$, which can be linked to ${\bf{s}}_i$ 
obtained from the Burgers vectors. Therefore,
knowledge of the moir\'e vectors ${\bf{m}}_i$ and the Burgers vectors across the dislocations associated with ${\bf{m}}_i$ is sufficient to solve for the displacement gradient matrix using Eq.~\ref{eq:solve-d}.
As the moir\'e vectors and Burgers vectors can be measured from experimental dark field and diffraction images, 
it is possible to obtain a map of $\overline{d}$ from the information provided by DF TEM images, as we will discuss in Sec.~\ref{sec:strainmapping}. 
	
	
	

\subsection{Configuration space}
Although a torus is topologically nontrivial, a small loop around the AA site (${\bf u} = 0$) on the torus can be contracted to a point and thus does not inherit any nontrivial topological properties from the torus. The topological defects associated with winding around the holes of the torus are the dislocations~\cite{Mermin}, but this description alone provides no constraint on the manner in which the dislocations meet at the nodes. In the case of a graphene moir\'e superlattice, the three dislocation lines with different Burgers vector, colored red($R$), green($G$), and blue($B$), converge at the AA defect and diverge out again. What we need is a proper mathematical description of the topological nature of the AA nodes consistent with the experimental findings. 

\begin{figure}
\includegraphics[width=\linewidth]{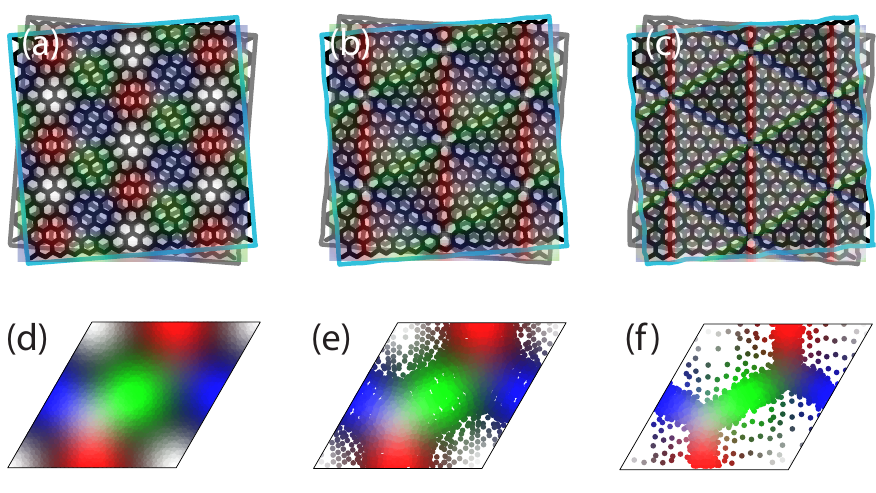}
\caption{\label{fig:inc-relax} (a-c) schematic of real space arrangement for (a) unrelaxed moir\'e structure, (b) partially relaxed moir\'e structure, (c) fully relaxed moir\'e structure. Colors denote region in configuration space corresponding to local stacking order.
(d-f) distribution in configuration space of local stacking order sampled at each plaquette of the lattice for (d) unrelaxed moir\'e structure, (e) partially relaxed moir\'e structure, (f) fully relaxed moir\'e structure. 
}
\end{figure}

 We start by investigating the distribution in configuration space of the order parameter for unrelaxed and relaxed moir\'e systems. Fig. \ref{fig:inc-relax}(a) shows an unrelaxed twisted bilayer of a honeycomb lattice, and Fig. \ref{fig:inc-relax}(d) shows the configuration space, colored via a scheme that is used consistently throughout this work. A Gaussian intensity distribution of a chosen color is centered around each key point in configuration space. The region of configuration space centered around the AA point is colored white, and those centered around AB (BA) are dark gray (light gray). Red, green, and blue regions are placed on the midpoint of the three equidistant shortest paths between AB and BA, two of which require crossing the unit cell boundaries. 
 The coloring in real space in Fig.~\ref{fig:inc-relax}(a-c) is determined for each point, by determining the local shift vector, finding it as a point in configuration space, and adopting the color corresponding to that point. 
 
 Fig.~\ref{fig:inc-relax}(b-c) depict real space structures that have been modulated by a periodic lattice distortion to mimic relaxation, with two different amplitudes. As the amplitude of relaxation is increased from Fig.~\ref{fig:inc-relax}(a) to (b) to (c), light or dark gray regions, corresponding to nearly AB or BA stacking, take up increasing areas in real space, while AA regions shrink, and red, green, and blue regions evolve into lines. Because the red, green, and blue color tell us which path on the torus was used to get between BA and AB, i.e. $\Delta \v u$, the color tells us the Burgers vector of that line. The corresponding distributions in configuration space are shown in Fig.~\ref{fig:inc-relax}(d-f). As relaxation strength increases, decreasing point density around the AA configuration reveals that fewer points in real space correspond to AA stacking, while configurations on the red, green, and blue lines, and especially AB and BA sites, become more numerous. 
Thus, the atomic lattice relaxation process in real space can be viewed as emptying out most of configuration space and populating only the AB and BA points and dislocation lines connecting them. 

\begin{figure}
\includegraphics[width=\linewidth]{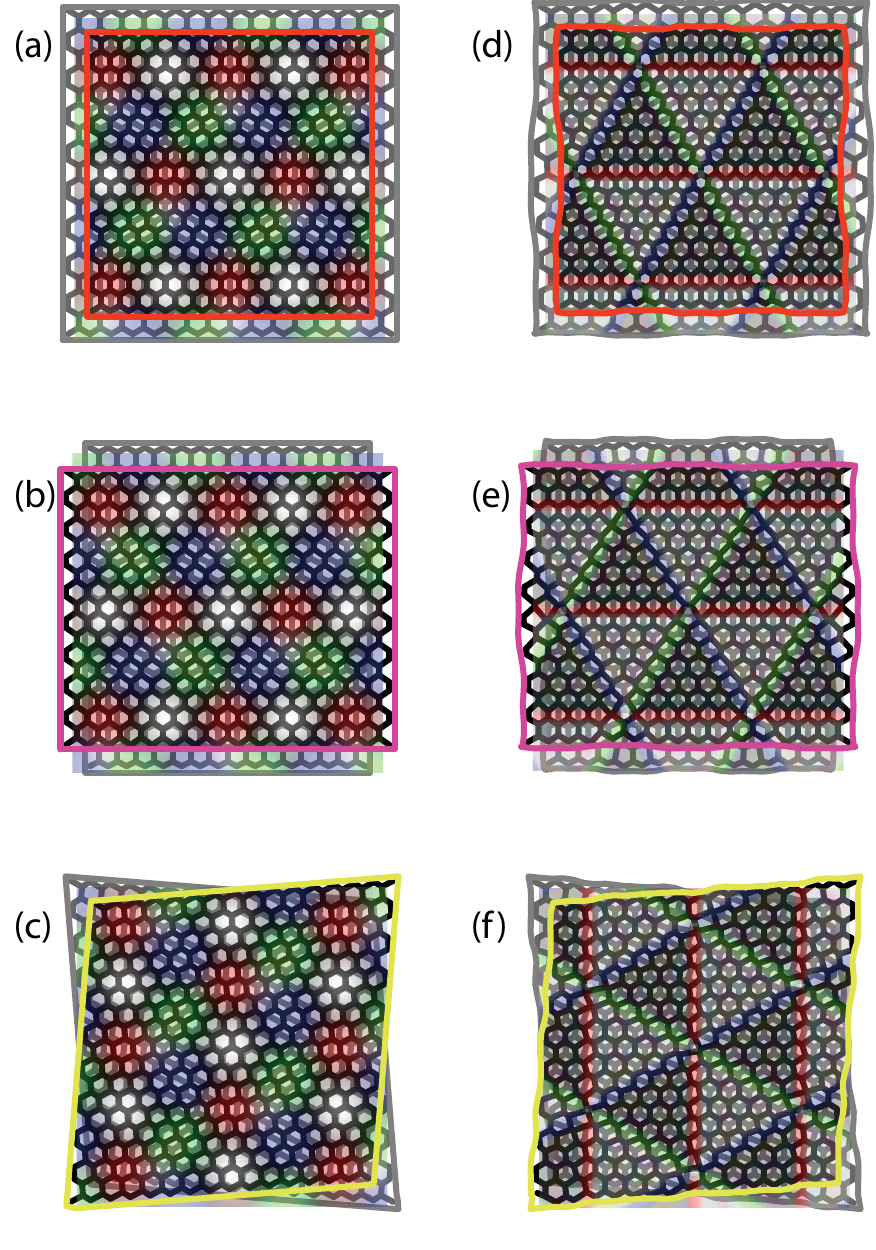} 
\caption{\label{fig:straintypes} Schematics of real space structure for the three strain components. Moir\'e from (a)  isotropic scaling, (b) uniaxial strain (c) shear strain. Relaxed moir\'e from (d) isotropic scaling, (e) uniaxial strain, (f) shear strain. Colors correspond to configurations, as defined in Fig.~\ref{fig:inc-relax}. 
}
\end{figure}

A similar emptying of AA and concentration at AB, BA, and the colored lines between AB and BA, occurs for the three strain types: isotropic, uniaxial, and shear. Fig.~\ref{fig:straintypes}(a-c) show the configurations formed by ${\bf{x}_t}=(1+\overline{d}) {\bf{x}_b}$ with a spatially constant displacement gradient matrix $\overline{d}$, corresponding to only one nonzero component $\alpha, \beta,$ or $\gamma$ in Eq.~\ref{eq:defd}. 
Fig.~\ref{fig:straintypes}(d-f) show corresponding structures after applying the modulation function to mimic atomic scale relaxation. We note that the order and orientation of red, green, and blue lines differs in real space for each strain component, as well as twist. In this sense, the colors (encoding local order parameter) provide information about which strain components are present.

\section{Theory of 
dislocation network nodes in moir\'e superlattices}
\label{sec:3} 
\begin{figure}
\includegraphics[width=\linewidth]{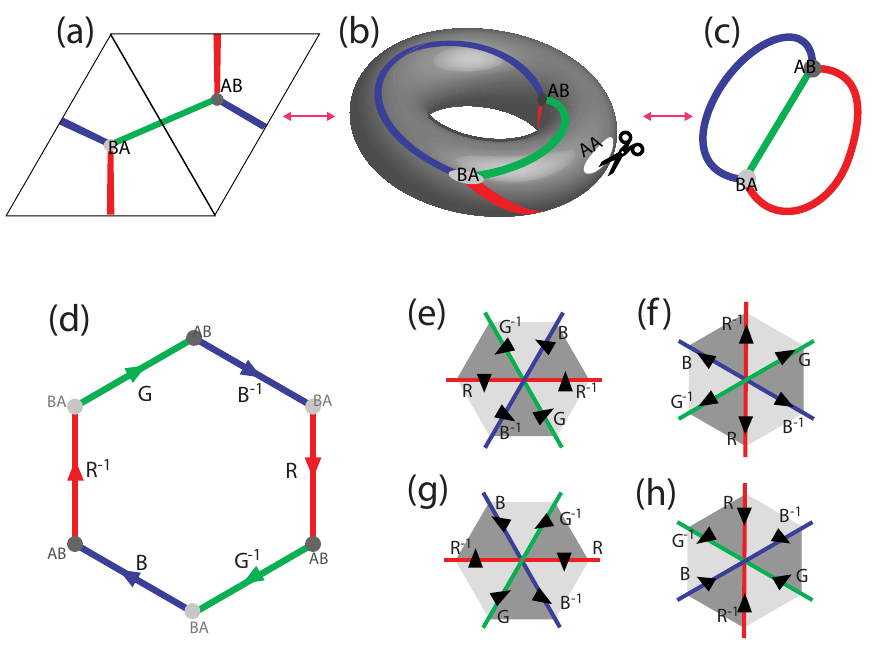}
\caption{\label{fig:topology} Energy 
landscape in the configuration space torus determines the topology of the network, which can be equivalently viewed (a) on the unit cell, (b) as a torus with a puncture (AA configuration is removed from space), or (c) as theta space. (d) Using the hexagonal unit cell, it can be seen that clockwise or counterclockwise paths around an AA point determine the order in which $R$, $G$ and $B$ elements are encountered. (e-h) Real space arrangement of dislocations corresponding to clockwise (e, f) and counterclockwise (g, h) paths in configuration space. Each domain wall is colored and labeled based on the R, G or B move in configuration space. The direction of the configuration space move, equivalent to the Burgers vector, is shown by the black arrows. Comparison with Fig. \ref{fig:straintypes} identifies the structures as generated from: (e) isotropic, (f) twist, (g) uniaxial, (h) shear displacement.}
\end{figure}

As emphasized previously, the proper order parameter space for a moir\'e superlattice and the various network structures observed in it is the space of shift vectors ${\bf u}$ with the boundary conditions of a torus and one point ${\bf u} = 0$ removed due to energy considerations (and the ensuing atomic relaxation), as shown in Fig. \ref{fig:topology}(b). 
But the TEM images shown in Fig.~\ref{fig:df} suggest an even more constrained space for the order parameter.  For moir\'e regions with a sufficiently large moir\'e length scale, the order parameter is locked to either the AB (${\bf u}_{AB}$) or BA (${\bf u}_{BA}$) point in the configuration space. There are three equivalent ways to make a transition from BA to AB stacking orders, designated by red, blue, and green arrows in Fig.~\ref{fig:topology}(d) with corresponding label $R$, $G$, and $B$. The AB to BA transition is accomplished by their inverses, shown as $R^{-1}$, $G^{-1}$ and $B^{-1}$ in Fig. \ref{fig:topology}(d). Since the ${\bf u}_{AB}$, ${\bf u}_{BA}$ and the three $RGB$ lines connecting the two points span the entirety of the relevant order parameters, one can ``gouge out" the unnecessary portions of the punctured torus. The result is the theta space shown in Fig. \ref{fig:topology}(c). This is the relevant configuration on which to make a proper definition of vorticity, not the circle ($S^1$) where the usual homotopic classification of vorticity takes place~\cite{Mermin}. 

Before developing the formal theory of vorticity in the next subsection we complete the phenomenological classification of possible vortex patterns around the AA node. According to the TEM data, a path around a single AA node in real space simultaneously implies encircling the AA spot in configuration space by ${\bf u}$. Closed paths in real space that encircle a node of the dislocation network now correspond to non-contractible paths in configuration space. The vortex winding number $w$ can be intuitively defined as $+1$ or $-1$ if the configuration space loop cycles the same or opposite direction as the real space loop, respectively (Fig. \ref{fig:topology}(d)). Recalling that the red, green, and blue paths in configuration space correspond to Burgers vectors of dislocations in real space, there are four distinct orderings of Burgers vectors in real space, two corresponding to vortices ($w=+1$, Fig. \ref{fig:topology}(e-f)), and two corresponding to antivortices ($w=-1$, Fig. \ref{fig:topology}(g-h)). Comparing the arrangement of Burgers vectors around the loop with the configurations in Fig.~\ref{fig:straintypes}, it can be concluded that twist and isotropic strain produce vortex-type defects at the nodes, and uniaxial or shear strain produce antivortex-type defects at the nodes.


\subsection{Algebraic formulation of vorticity}

\begin{figure}[!ht]
\includegraphics[width=0.5\textwidth]{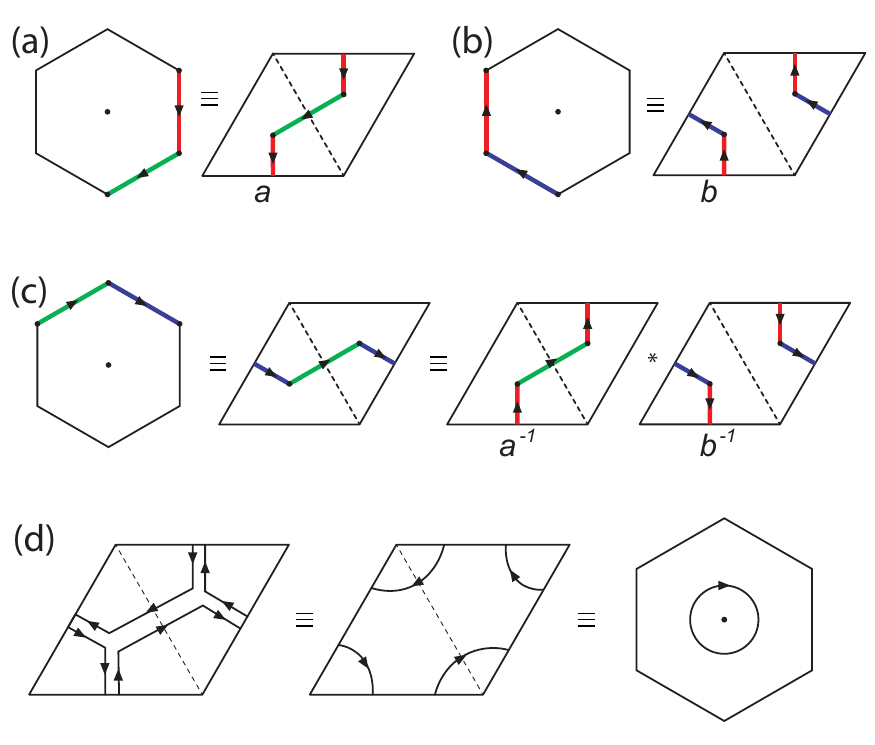}
\caption{(a) The $RG^{-1}$ move in the hexagonal zone scheme of the configuration space is translated into a non-contractible loop in the parallelogram zone scheme. It is labeled by $a$. (b) The $BR^{-1}$ move in the hexagonal zone scheme of the configuration space is translated into a second non-contractible loop in the parallelogram scheme. It is labeled by $b$. (c) The $GB^{-1}$ move in the hexagonal zone scheme is translated into a third non-contractible loop that can be decomposed as a product of the previous two moves, $a^{-1}b^{-1}$. (d) The $ab a^{-1} b^{-1}$-loop is equivalent to a circle round the point $\v u =0$. It is a non-contractible loop due to the high-energy barrier for the $\v u=0$ configuration.}
\label{fig:ab-moves}
\end{figure}

To understand the algebraic structure of paths in the theta space (Fig. \ref{fig:topology}(c)), it is useful to consider the same path through configuration space in both the unit-cells shown in Fig. \ref{fig:shift}(e).
Let us consider the full encirclement of the AA node shown in Fig. \ref{fig:topology}(d).
We begin by observing that the BA $\rightarrow$ AB $\rightarrow$ BA transition through the $R$ and $G^{-1}$ dislocations in Fig. \ref{fig:topology}(d) becomes, in the parallelogram scheme shown in Fig. \ref{fig:ab-moves}(a), a non-contractible loop about the torus' periodic boundary conditions. The other BA $\rightarrow$ AB $\rightarrow$ BA transition through the $B$ and $R^{-1}$ arrows in the hexagonal scheme becomes another non-contractible loop in the parallelogram unit cell shown in Fig. \ref{fig:ab-moves}(b). We label the first and the second transitions as $a$ and $b$, which will soon be identified with two generators of the free group $F_2$~\cite{free-group,free-nonabelian-group}. The third BA $\rightarrow$ AB $\rightarrow$ BA move through $G$ and $B^{-1}$ arrows of the hexagon can be decomposed as the inverse of $a$ followed by the inverse of $b$, or $a^{-1} b^{-1}$ [Fig. \ref{fig:ab-moves}(c)]. A complete loop around the edges of the hexagonal unit cell is equivalent to the algebraic operation $a b a^{-1} b^{-1} \equiv [a,b]$. Our convention is to perform the operation appearing on the left side of the product first.

The commutator $[a,b]$ in the language of the free group with two generators $a, b$ represents a ``vortex" centered about the AA defect. The vorticity 
of this topological defect can naturally defined as the commutator $[a, b]$ (more detailed discussion in Appendix \ref{sec:appendix-1}). Similar to the conventional vortex defined in $S^1$, this vorticity defined for the AA node is non-trivial in the sense that it is not contractible to an identity, as illustrated in Fig. \ref{fig:ab-moves}(d). After cancelling out the paths that are traversed both ways, the overall path for $[a,b]$ becomes equivalent to four partial loops around the four corners of the parallelogram, equal to a full loop round $\v u =0$. On a torus such a loop can be contracted to zero and become trivial, but not for a punctured torus. The anti-vortex has the algebraic representation $[b,a]=b a b^{-1}a^{-1} = [a,b]^{-1}$. Geometrically, this amounts to starting from the same BA point on the upper left corner of the hexagonal unit cell in Fig. \ref{fig:topology}(d) and making a complete counter-clockwise loop. Appendix \ref{sec:appendix-1} gives a more complete account of the vortex structures in the language of free groups with two generators $a, b$. One can find group-theoretic representations for vortex dipoles (vortex + antivortex) and vortex quadrupoles (two vortices and two antivortices) as well. 


\subsection{RGB formulation of vorticity}
As we described in Section \ref{sec:2}A, dark field TEM imaging can identify the dislocation lines with given Burgers vectors and the AB and BA domains separated by them. Each dislocation line converging on a given AA node can then be color-coded  as $R$, $G$, $B$ or one of their inverses $R^{-1}$, $G^{-1}, B^{-1}$ considering Burgers vector and the neighboring AB/BA domains in the TEM measurement. The free group language of the previous subsection gives a mathematically complete account of the vortex and antivortex structures, but it is helpful to translate the same statement to the more tangible and experimentally measurable $RGB$ scheme according to 
\ba a \leftrightarrow R G^{-1} , ~~ b \leftrightarrow B R^{-1} ~~ ba \leftrightarrow B G^{-1}.  \label{eq:ab-RGB} \ea
By direct substitution we obtain the commutator 
\begin{align} [a,b] = R G^{-1} B R^{-1} G B^{-1},  ~~~ ({\rm vortex}) \label{eq:RGB-vortex} \end{align} 
which is a product of transition vectors over the six domain walls in succession. In the same scheme we have the antivortex commutator 
\begin{align} [b,a] = B G^{-1}  R B^{-1} G R^{-1} ~~~ ({\rm antivortex}). \label{eq:RGB-antivortex} \end{align} 
Any cyclic permutation of the six letters gives rise to the equivalent vortex or antivortex. 

While the sign of the exponent in $R$, $G$, and $B$ operators can be obtained considering the order of the neighboring AB and BA domains by combining the DF TEM images with the first and second order Bragg peaks as shown in Fig.~\ref{fig:df}, there is a simpler scheme to assign the sign of the operators, considering AB/BA domains are always complementary. For example,
if the six dislocation lines converging on an AA node appear, for instance, in the order of $RGBRGB$ while going clockwise 
around it, it ought to be interpreted as $R G^{-1} B R^{-1} G B^{-1}$ given in Eq. (\ref{eq:RGB-vortex}) and classified as a vortex. If the colors appear as $RBGRBG$, it is an anti-vortex according to Eq. (\ref{eq:RGB-antivortex}). One only needs to keep in mind that the sequence of colors is to be understood as one color letter followed by the inverse of another color letter, and vice versa. Generalizations of the RGB scheme to vortex-antivortex dipole and/or vortex quadrupole structure are discussed in the Appendix \ref{sec:appendix-2}. 


\section{Experimental observation of antivortices and strain} 
\label{sec:4} 

\subsection{Detection of antivortices}
While moir\'e patterns and commensurated domain systems with vortex-type nodes have been studied extensively, those with antivortex-type nodes have not been demonstrated. We postulate this is due to the energy required to maintain sufficient strain, whereas twist and lattice constant mismatch can create vortex-type moir\'e without global strain. Nonetheless, we observe a line of antivortex nodes along a boundary of non-uniformly strained moir\'e superlattice.

\begin{figure}
\includegraphics[width=\linewidth]{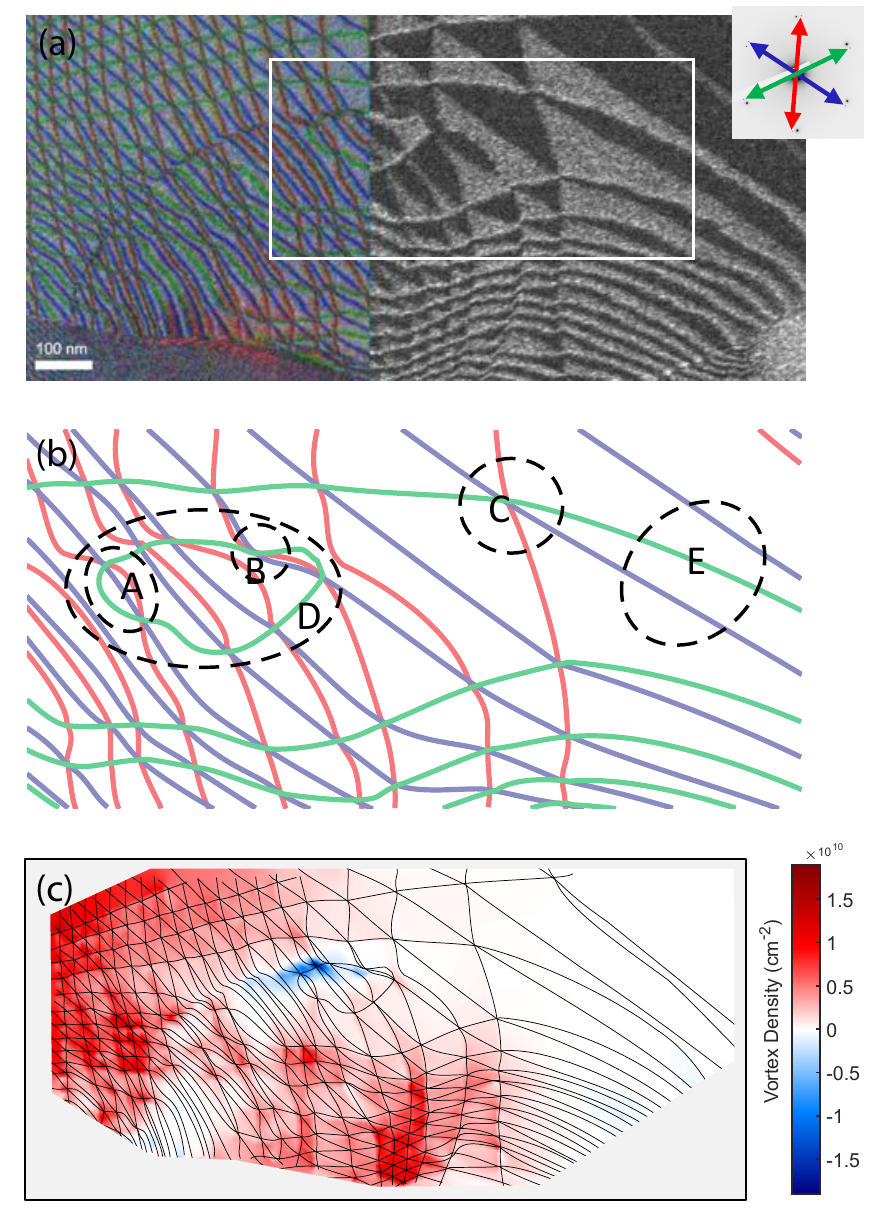}
\caption{\label{fig:av-data} a) Dark field image of twisted bilayer graphene containing antivortices along a bubble edge. Left region shows colored second order image, right region shows first order image. Inset: Burgers vector directions corresponding to dislocation colors, superimposed on diffraction pattern. 
b) Zoom in on tracing of region of white box in 1a. Loops are drawn and topological number of each loop is counted. A) Vortex-antivortex pair, $w=0$. B) Antivortex, $w=-1$. C) Vortex, $w=+1$. D) Closed-loop dislocation, $w=0$. E) Linear domains, $w=0$.
c) Vortex density map computed from interpolated displacement gradient matrix of image (a)
}
\end{figure}

Fig.~\ref{fig:av-data}(a) shows combined DF TEM images of a twisted bilayer graphene sample that contains a $\sim$1~$\mu$m sized bubble formed underneath the sample. Near the boundary of the bubble, non-uniform relative strain builds up in the moir\'e superlattice, which in turn creates the various strain components discussed in Eq.~\ref{eq:defd}.
The antivortices (vortices) can be identified by the $RBG$ ($RGB$) order in which the dislocations occur in a clockwise loop. The antivortices, which form along the top edge of a closed-loop dislocation, are each capable of annihilating with a nearby vortex, keeping the net winding number $w$ constant within a fixed-boundary region. In Fig. \ref{fig:av-data}(b), loop A surrounds a vortex-antivortex pair, with $w=0$. Loop D, which surrounds the entire closed-loop dislocation, also has net $w=0$ as the entire feature could annihilate if the local strain were removed, in which case it would become similar to loop E. When circling a single antivortex (B) or vortex (C), the winding number is nonzero.

\subsection{Strain mapping}
\label{sec:strainmapping}
Existence of antivortices is a measure of the fact that anisotropic strains (uniaxial and shear) are dominating over the isotropic and twist components. Anisotropic strains alter the band structure and can produce pseudomagnetic fields~\cite{PseudoGeim}. We can quantify the various strain components from the displacement gradient matrix. Computation of the displacement gradient matrix from a DF TEM image is discussed in Appendix~\ref{sec:appendix-imageprocessing}. In brief, one component of the order parameter is known at every colored line, and an elastic model is used to interpolate in between, obtaining an estimate of the order parameter in the continuum. The displacement gradient matrix can be obtained by differentiating.

The density of vortices or antivortices is computed in Fig. \ref{fig:av-data}(c), by taking the determinant of the displacement gradient matrix, $|\overline{d}|$, which by the definition in Eq.~\ref{eq:defd} is equal to $(\alpha^2+\theta^2) - (\beta^2 + \gamma^2)$ in terms of the strain and twist components. If $|\overline{d}|>0$, vortices are present and if $|\overline{d}|<0,$ antivortices are present (see Appendix~\ref{sec:appendix-imageprocessing}). Thus, if the isotropic components (twist and isotropic scaling) outweigh the anisotropic components (shear and uniaxial strain), vortices form, and if the opposite, antivortices form. 

\begin{figure}
\includegraphics[width=\linewidth]{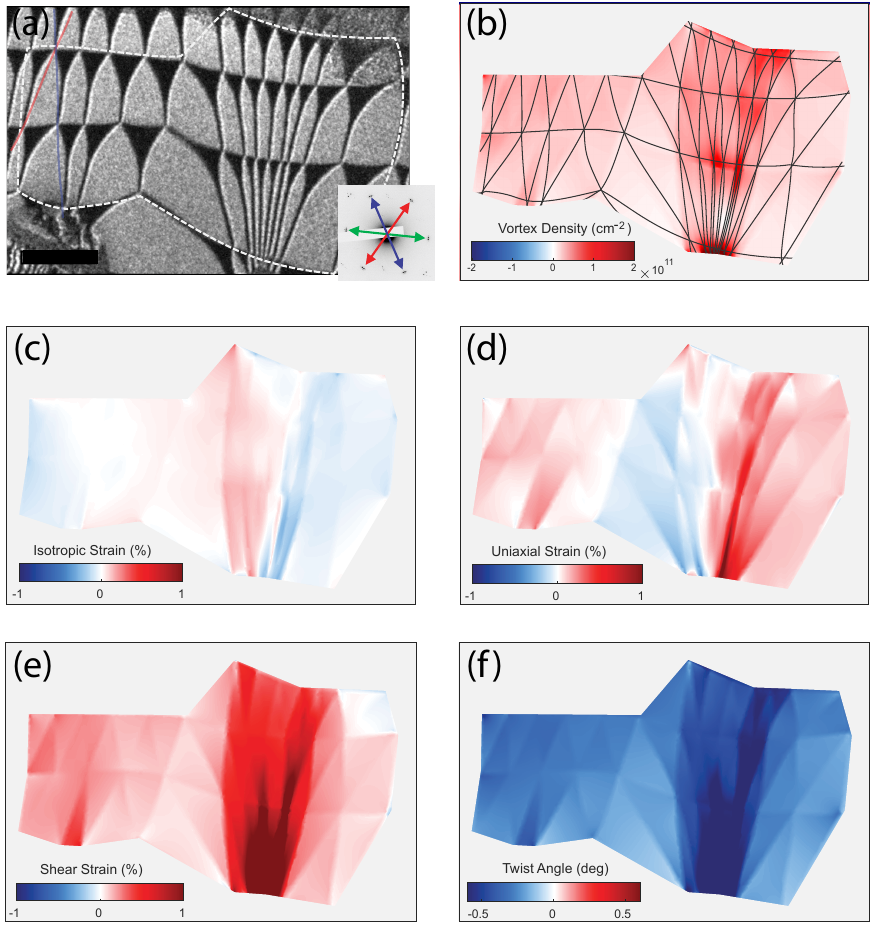}
\caption{\label{fig:strainmaps} a) Dark field image of WSe$_2$/MoSe$_2$ heterobilayer, including domains larger than the lattice-mismatch limit. Inset: Diffraction pattern from a nearby region and Burgers vectors. 
(b) Vortex density map.
(c) Isotropic strain map showing average mismatch lower than the 0.3\% expected from the intrinsic lattice mismatch.
(d) Uniaxial strain map, showing opposite sign strain when domains are slanted left vs right.
(e) Shear strain, showing magnitudes around 1\% in the highly elongated domains.
(f) Twist map, showing twist as the largest contributor to the moir\'e pattern.
}
\end{figure}

We further use our estimated displacement gradient matrix to create strain maps of the three strain components, plus twist. Note that the moir\'e pattern only provides information on the heterostrain, or difference in strain between the two lattices. Unlike other methods to estimate heterostrain from the spatial structure of a moir\'e pattern~\cite{DorriStrain}, this DF TEM method includes knowledge of the lattice orientation and Burgers vector information, avoiding the need to make additional assumptions. Still, the need to interpolate within the moir\'e cell means that information smaller than the moir\'e scale is not deterministic from the data. In Fig. \ref{fig:strainmaps}, strain maps are shown for a heterostructure of MoSe$_2$ and WSe$_2$. The intrinsic lattice constant mismatch of 0.3\% should show up in the isotropic component. However, isotropic mismatch smaller than 0.3\% is measured, indicating that the lattice globally strains to achieve closer to epitaxial matching, whereas global lattice mismatch is often assumed to be fixed when calculating moir\'e lengths. This type of strain mapping can reveal useful information about the phenomenology of moir\'e materials. 

\section{Conclusion}
\label{sec:5}

\begin{figure}[!ht]
\includegraphics[width=0.5\textwidth]{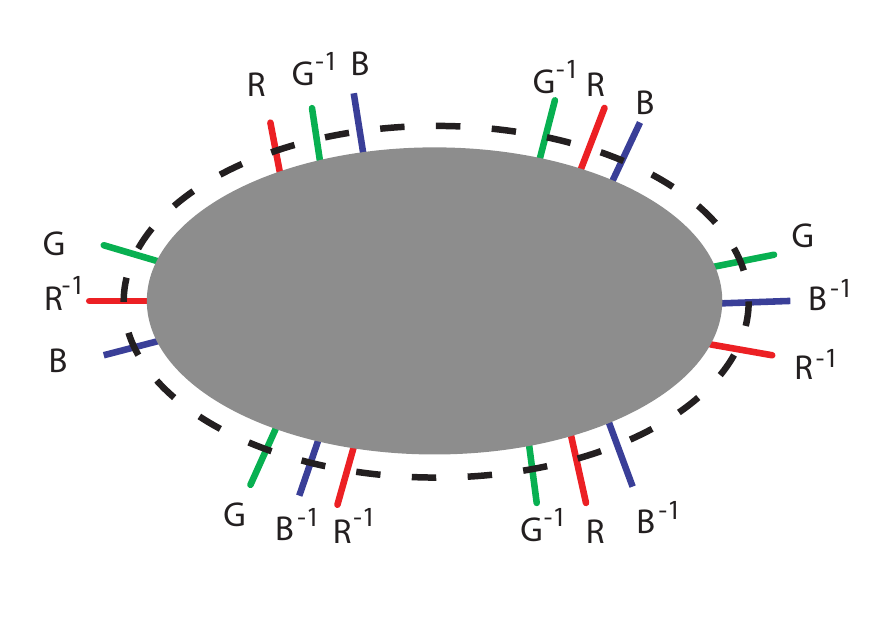}
\caption{General configuration of RGB lines extending from an arbitrary circle drawn on an experimentally TEM image. By writing out the RGB letters along the circumference of the circle, one can read off the total vorticity, number of dipoles, etc. contained in the circle.}
\label{fig:general-RGB-scheme}
\end{figure}

In conclusion, we have presented a general and rigorous approach to describing the topology of nodes formed in moir\'e materials. Vortex and antivortex are described as the commutator $[a,b]$ and its inverse $[b,a]$ of the free group $F_2$ on generators $a$ and $b$. The two generators have an intuitive geometric interpretation as two distinct ways by which to make a transition from the AB to BA stacking order, and then back to AB. High-quality TEM measurements are then utilized to represent the abstract generators in terms of colors of domain walls, leading to a dictionary by which to infer the vortex content inside a given boundary.
This dictionary relies upon the order in which the colored domain walls cross the boundary. The idea is schematically illustrated in Fig. \ref{fig:general-RGB-scheme}. 

We discover an antivortex-type node and present a DF TEM based method for characterizing the type of node and strain field, which does not rely on the usual assumption that the dominant component creating the moir\'e pattern is twist. This opens the door for the design and characterization of moir\'e materials based on anisotropic strain fields. 

\section{Acknowledgement}
\label{sec:Ack} 
We thank Frans Spaepen, David Nelson, S.-W. Cheong for important discussions. P.K. acknowledges support from ARO MURI (W911NF-21-2-0147). R.E. acknowledges support from NSF DMR-1922172 for TEM analysis. H.Y. acknowledges the support by the National Research Foundation (NRF) grant funded by the Korean government (MSIT) (No.2021R1C1C1010924). S.C. acknowledges support from NSF Grant No. OIA-1921199. P.C. acknowledges support from NSF DMS Award No. 1819220.  ML acknowledges support from NSF DMREF Award No. 1922165. J.H. was supported by  NRF-2019R1A6A1A10073079 and by EPIQS Moore theory centers at MIT and Harvard. He acknowledges the help of Manhyung Han for drawing the figures in the Appendix and the discussion with Harry Baik on free group.

\appendix 
\section{Vortex algebra}
\label{sec:appendix-1}

We have avoided explicit use of the language of group theory in the main body of the paper. The two generators $a, b$ and their commutators $[a,b]$, $[b,a]$ were introduced through physical motivation. Here we provide more in-depth discussion and generalizations based on the theory of the free group.

The fundamental group of the punctured torus is $F=F_2$, the free group on two generators $a, b$~\cite{free-group,free-nonabelian-group}. The $a$ and $b$ generators correspond to the two independent ways in which one can encircle the torus. In an ordinary torus the two operations $a$ and $b$ do commute (Abelian), and the only elements of the fundamental group of the torus are $a^m b^n$, which count the number of loops in both directions. For a punctured torus such commutativity is lost, and consequently the group structure becomes non-Abelian. 

Elements of the free group $F_2$ consist of every conceivable sequence of ``letters" such as $a b a a b b b a a \cdots$ called ``words". Keep in mind that both letters $a$ and $b$ have specific geometric moves associated with them. At this point it is helpful to go over well-established theorems in free groups to guide our thinking.  

We now use $F$ to denote the original free group $F_2$. Given two elements $x, y \in F$, the commutator is denoted $[x,y] \equiv x y x^{-1} y^{-1}$. The ``lower central series" of the free group can be defined as follows. One begins with $F^1=F$ which is the original free group, then $F^2 = [F, F]$ is the subgroup of $F$ consisting of all commutators $[x,y]$ and their products, i.e., all elements of the form
\ba [x_1 , y_1 ] [x_2 , y_2 ] \cdots [x_n , y_n ] , ~~ (x_i, y_i \in F) . \nonumber \ea
The subgroup $F^2$ is also a normal subgroup, meaning that the quotient space $F^1 / F^2$ is a group. 

Now one can proceed inductively and define 
\begin{align} F^n = [ F, F^{n-1}], \nonumber \end{align}
the subgroup generated by all elements of the form $[x,y]$ where $x\in F$ and $y \in F^{n-1}$. It is easy to check from the definition that
\ba F^1 \supset F^2 \supset F^3 \supset F^4 \cdots \nonumber \ea
Much like the study of van der Waals materials, such ``filtration" gives a nice way to study a free group structure `one layer at a time'!

Some facts that are worth noting about the lower central series are summarized:

\begin{enumerate}
\item Any element $f_1$ of $F^1 =F$ can be uniquely written $a^{m_1} b^{n_1} f_2$ with $n_1, m_1 \in \mathbb{Z}$ and $f_2 \in F^2$. 

\item Any element $f_2$ of $F^2$ can be uniquely written $[a,b]^{m_2} f_3$ where $m_2 \in \mathbb{Z}$ and $f_3 \in F^3$. 

\item Any element $f_3$ of $F^3$ can be uniquely written $[a,[a,b]]^{m_3} [b,[a,b]]^{n_3} f_4$ with $m_3, n_3 \in \mathbb{Z}$ and $f_4 \in F^4$. 

\item Any element $f_4$ of $F^4$ can be uniquely written $ ( [a, [a, [a,b]])^{m_4} ([a,[b,[a,b]]])^{n_4} ([b,[b,[a,b]]])^{p_4} f_5$ with $m_4 , n_4, p_4 \in \mathbb{Z}$ and $f_5 \in F^5$. 
\end{enumerate}
By putting all of the above statements together, one sees that any element $f$ in the free group can be uniquely written as
\begin{widetext}
\ba f = a^{m_1} b^{n_1} [a,b]^{m_2} [a, [a,b]]^{m_3} [b,[a,b]]^{n_3} ( [a, [a, [a,b]])^{m_4} ([a,[b,[a,b]]])^{n_4} ([b,[b,[a,b]]])^{p_4} f_5 \label{eq:F2-elements} \ea
\end{widetext}
and so on. In general, each $F^k$ is a normal subgroup, and $F^k/F^{k+1} \simeq  \mathbb{Z}^{r_k}$, meaning the quotient group is isomorphic to a product of $r_k$ integer groups $\mathbb{Z} \times \cdots \times \mathbb{Z}$. The number of generators is $r_k$ for a given quotient group $F^k / F^{k+1}$. Although the free group itself is non-Abelian, the quotient group $F^k / F^{k+1}$ is Abelian, characterized by a set of $r_k$ integers. These integers then go on to play the role of topological quantum numbers in physical contexts. 

\begin{figure}[!ht]
\includegraphics[width=0.4\textwidth]{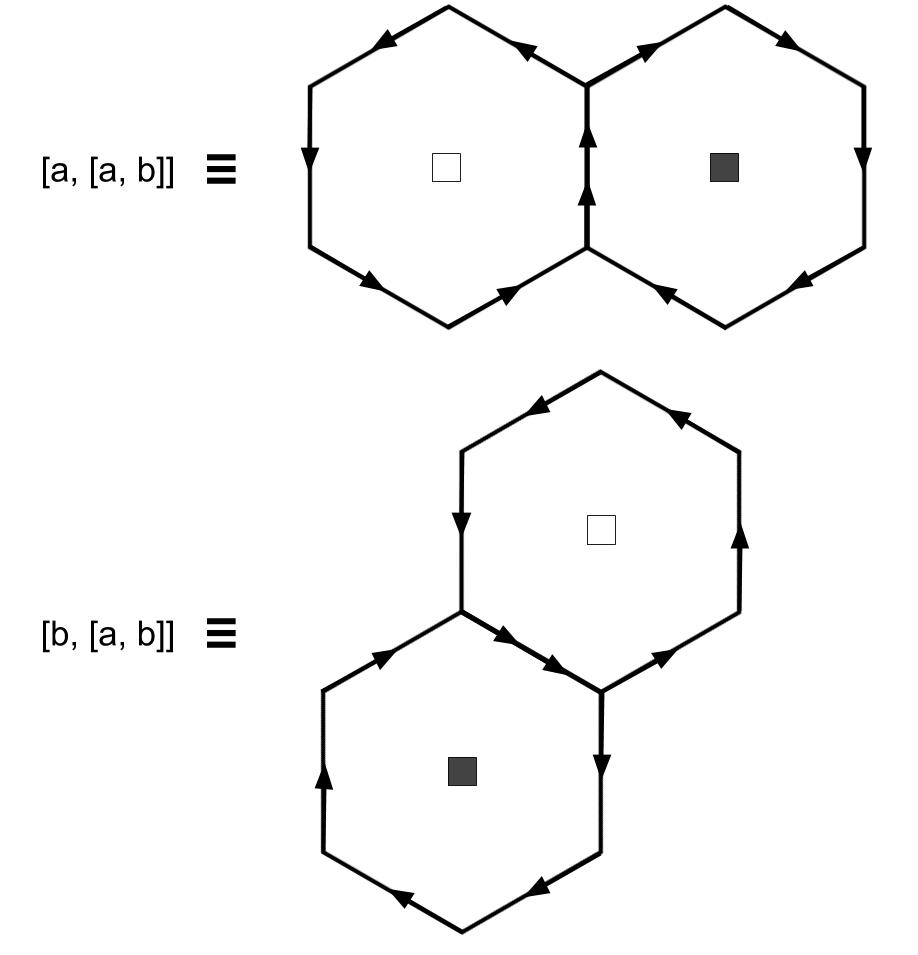}
\caption{Graphical representation of the double commutators $[a,[a,b]], [b,[a,b]]$. They give vortex dipoles oriented along the two directions of the triangular lattice. Filled (empty) square is a vortex (anti-vortex). }
\label{fig:appendix-1}
\end{figure}

Elements of the free group (\ref{eq:F2-elements}) for which $m_1 = n_1 =0$ refer to closed loops in real-space graphical representation. It is clear that these are the only elements of $F$ that we are interested in. Elements for which $f_5 = e$ (an identity) and $m_3 = n_3 = m_4 = n_4 = p_4 = 0$ are $f = [a, b]^{m_2}$ with nonzero $m_2$. These are the elements of the quotient group $F^2 / F^3$ and represent the vortices  ($m_2 >0$) and antivortices $(m_2 < 0)$ in physical contexts. 

To consider higher-order topological defects, consider elements for which $f_5 = e$ and all integers in Eq. (\ref{eq:F2-elements}) equal to zero except $(m_3, n_3)$: 
\begin{align} f = [a,[a,b]]^{m_3}  [b,[a,b]]^{n_3}. \end{align} 
Pictorial representations for the double commutators $[a,[a,b]], [b,[a,b]]$ are easily obtained by tracing out paths according to definitions of $a$ and $b$ given in Fig. \ref{fig:ab-moves}. We encourage readers to perform such exercises themselves and arrive at their graphical representations shown in Fig. \ref{fig:appendix-1}. They are precisely the graphical representation of vortex-antivortex pairs (vortex dipoles) lying along the two crystallographic directions of the triangular lattice.  

\begin{figure}[!ht]
\includegraphics[width=0.5\textwidth]{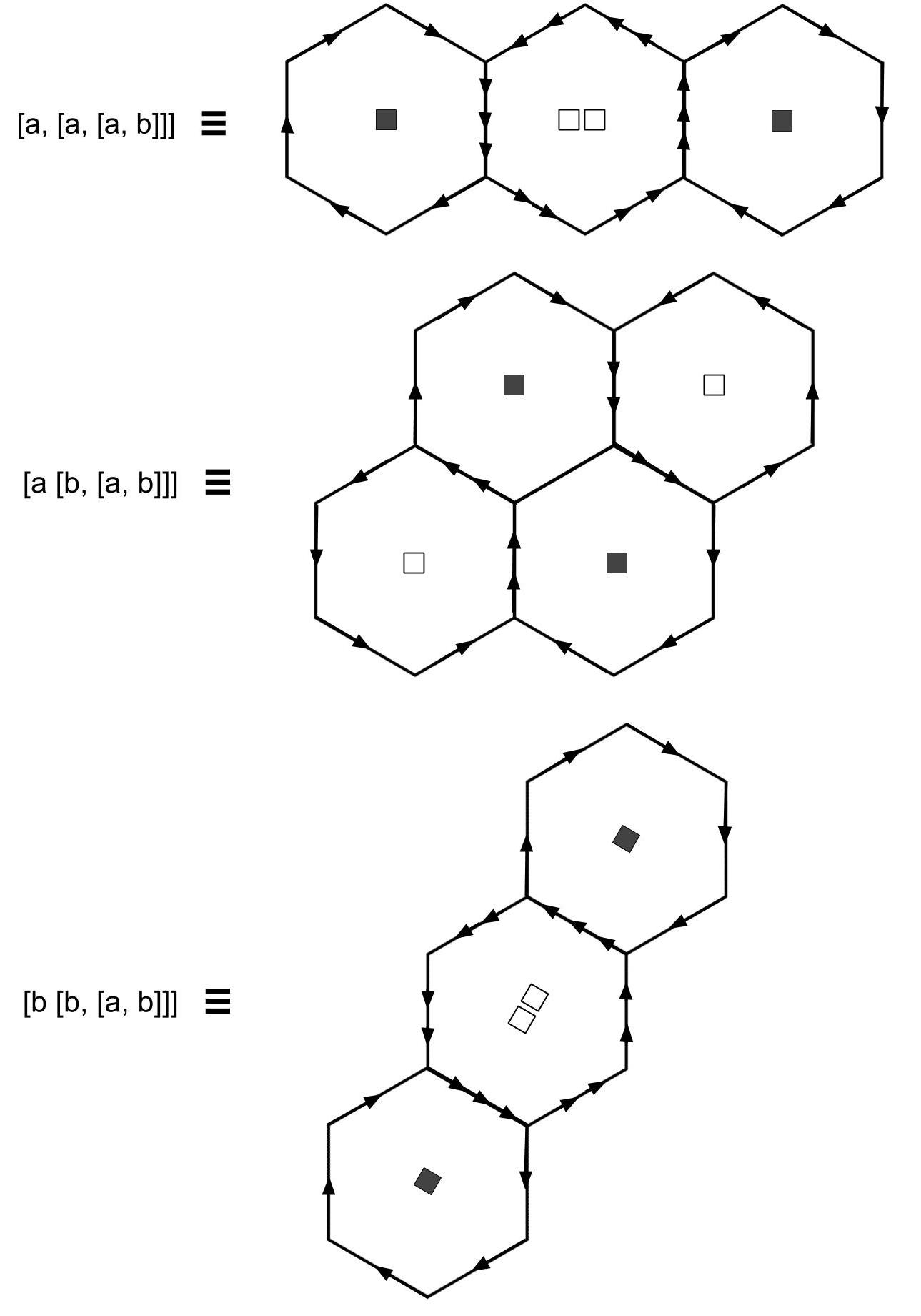}
\caption{Graphical representation of the three triple commutators $[a, [a, [a,b]], [a, [b,[a,b]]], [b,[b,[a,b]]]$ correspond to three different kinds of vortex quadrupoles in real space.}
\label{fig:appendix-2}
\end{figure}

Next in line is the description of vortex quadrupole structure as triple commutators. According to Eq. (\ref{eq:F2-elements}), there are only three generators of the quotient group $F^4/F^5 \simeq \mathbb{Z} \times \mathbb{Z} \times \mathbb{Z}$. How does one know there are only three generators at this level of filtration? 

There is a theorem that gives the number of generators $(r_k$) at each level $k$ through the formula
\ba g^k = \sum_{d | k } d \cdot r_d . \ea
In this formula the sum runs over all divisors $d$ of the given integer $k$. For a free group with only two generators we have $g=2$ on the left side of the equation. To see how the formula works with $g=2$, first set $k=1$ to find $2=r_1$. It means that the quotient group $F^1 /F^2$ has two generators, namely $a$ and $b$. At $k=2$ we have $2^2 = r_1 + 2 r_2 = 2 + 2 r_2$ or $r_2 = 1$, hence there is only one generator of $F^2 / F^{3}$ which is the commutator $[a,b]$. At $k=3$ we have $2^3 = r_1 + 3 r_3 = 2 + 3 r_3$, and $r_3 = 2$ is the number of generators for $F^3 / F^4$, namely $[a,[a,b]]$ and $[b,[a,b]]$. Finally, at $k=4$ we get $2^4 = r_1 + 2 r_2 + 4 r_4 = 2 + 2 + 4 r_4$, and $r_4 = 3$ is the number of generators of $F^4 /F^5$ given by $[a,[a,[a,b]]], [a,[b,[a,b]]], [b,[b,[a,b]]]$. It is an arduous, but fun exercise to draw the real-space paths corresponding to each of the triple commutator. The results are the three distinct vortex quadrupole configurations in real space shown in Fig. \ref{fig:appendix-2}. 


\section{RGB scheme for higher-order vortices}
\label{sec:appendix-2}

In Sec. \ref{sec:3} we discussed ways to characterize a vorticity in terms of the RGB color scheme. A similar RGB scheme to characterize various higher-order vortex structures can be developed. 

\begin{figure}[!ht]
\includegraphics[width=0.5\textwidth]{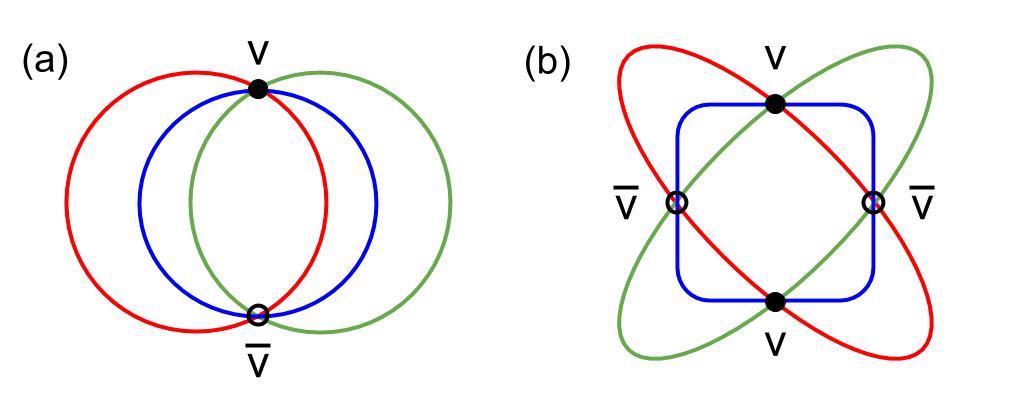}
\caption{a) RGB graphical representation of a vortex dipole with $v$(vortex, filled circle) and $\overline{v}$(anti-vortex, empty circle) sites. (b) Quadrupole configuration with alternating $v\overline{v}v \overline{v}$ cores.}
\label{fig:appendix-3}
\end{figure}

Fig. \ref{fig:appendix-3} shows the vortex dipole and quadrupole configurations in terms of intersecting RGB loops. A small circle drawn around each intersection can determine the vorticity of that point. For instance the filled (empty) circle round the top (bottom) intersection in Fig. \ref{fig:appendix-3}(a) reads the product of letters $R G^{-1} B R^{-1} G B^{-1}$ ($B G^{-1}  R B^{-1} G R^{-1}$) going counter-clockwise, corresponding to a vortex (an antivortex). Vortex quadrupole construction is done by having the three RGB loops intersect at four different points, as shown in Fig. \ref{fig:appendix-3}(b).  In both cases, a large circle drawn far away from the loops fails to cross any of the RGB lines. 

\begin{figure}[t]
\includegraphics[width=0.5\textwidth]{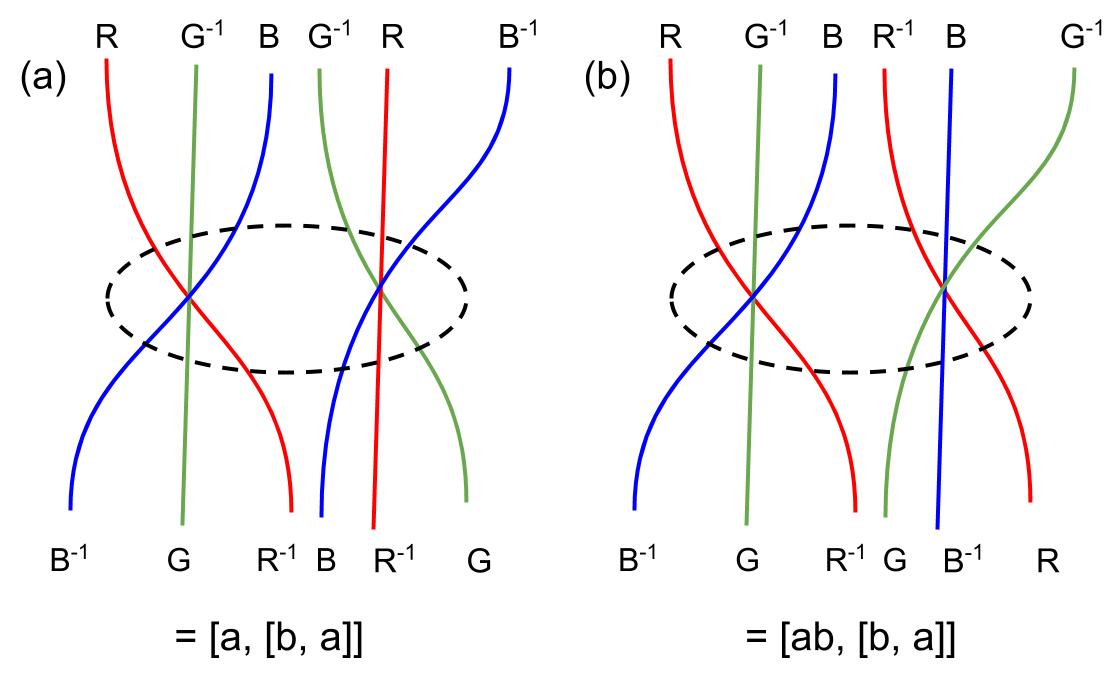}
\caption{(a) Extended vortex dipole configuration characterized by the letter sequence $RG^{-1}BG^{-1}RB^{-1}GR^{-1}BR^{-1}GB^{-1}$. After simple calculation, this becomes equivalent to $[a,[b,a]]$. (b) Another extended vortex dipole configuration, expressible as the double commutator $[ab, [b,a]]$.}
\label{fig:appendix-4}
\end{figure}

It is possible to construct examples of vortex dipole configurations with loops extending out to infinity (hence crossing an arbitrary large circle) as in Fig. \ref{fig:appendix-4}(a) and (b). The group elements assigned to each configuration can be calculated straightforwardly, leading to $[a, [b,a]]$ and $[ab, [b,a]]$ for the left and right configurations, respectively. 

This kind of scheme is applicable to experimental situations. Draw a large loop enclosing a given TEM image in the manner shown in Fig. \ref{fig:general-RGB-scheme}, then start counting the dislocation lines according to the $RGB$ scheme. The RGB-based words can be converted to the $ab$-letter scheme with the help of the dictionary given in Eq. (\ref{eq:ab-RGB}). The $ab$-based word can subsequently be converted to various commutators and higher-order commutators until it is cast in the general form given in Eq. (\ref{eq:F2-elements}), from which the total vorticity, dipole numbers, quadrupole numbers and so on can be read off.

\section{TEM image processing}
\label{sec:appendix-imageprocessing}

TEM DF images were taken on a JEOL 2010F microscope, with 80kV accelerating voltage. A $5\mu$m objective aperture was used to form the dark field images. RGB-colored composite images were formed using Adobe Photoshop blending modes from three distinct second order images of a given sample region.

To create strain maps in Fig.~\ref{fig:strainmaps}, we performed image processing using Python programming codes on images of red, blue and green lines, such as in Fig.~\ref{fig:av-data}(b), to prepare to interpolate shift vector values. Ideally an algorithm could be created to extract the lines straight from TEM images but we skip that step for now, and manually trace the red, green, and blue lines to create an ``ideal" (i.e. noiseless), but still raster, image. It is then necessary to split all lines into individual line segments and points of intersections, which we call nodes. First, using the connectedComponents function in Python's OpenCV library, nodes are found by searching for places where red, green, and blue pixels coincide. In a similar manner, pixels that belong to each red, green, or blue line are grouped into lists after dilating them to ensure continuity. We include each line in a dictionary that describes the color of the line and what nodes are part of it. Next, a circle centered around each node is removed from the image (set to R,G,B = 0,0,0), effectively breaking the lines into line segments. Again, the pixels of each segment are found and a dictionary is created for each line segment. The parent line of each segment is identified, as well as the nodes that are its endpoints.

Next, each node must be assigned three integer values corresponding to the coefficients of $\bf{a}_\text{R}, \bf{a}_\text{G}$ and $\bf{a}_\text{B}$, where $\bf{a_\text{R}}$ is the lattice vector associated with the red line, and so on. An origin node is picked arbitrarily and assigned the coefficients $(u_R,u_G,u_B)=(0,0,0)$. Then, a red segment adjacent to the starting node is chosen to reach a second node, which is assigned $(u_R,u_G,u_B) = (0, 1, -1)$. Now that a second node is assigned, the choices are not arbitrary because the direction of increase for each vector has been determined. Note that it is the case that $u_G+u_G+u_B=0$ at every node, because $\bf{a}_\text{R}+\bf{a}_\text{G}+\bf{a}_\text{B}=0$. We also know that $u_r$ is constant on red lines, $u_g$ on green lines, etc. From each node, we move to its neighboring nodes (those connected by a segment), and use these properties to fill in the rest of the coefficients. A nice property of this manner of assigning coefficients is it does not need to be told whether a given point is a vortex or an antivortex. Lastly, using the csap and scipy libraries, we fit B-splines to each line segment while forcing the spline to pass through the nodes of the segment, and then use the knots of the splines to generate a Gmsh mesh file.

The next part of the computation is done in the Julia programming language. The vector values of $\bf{a}_\text{R}$ and $\bf{a}_\text{G}$ (measured from a diffraction pattern) must be input ($\bf{a}_\text{B}$ can be found from the other two). Recall that the vectors in the image plane should appear in the order of RGB while going clockwise to correctly distinguish vortices and antivortices. Coefficients are then attributed to points on lines of the mesh and interpolated via an elastic model using the Gridap library~\cite{Badia2020} and its interface with Gmsh~\cite{Gmsh}. The elastic model applies a cost to a large derivative in the $u$-field, as well as a cost to deviating from the known values on the lines. The resulting $u$-field is differentiated to get strain components. The values are defined on a mesh that is small compared to the moir\'e length. For plotting, the mesh values are interpolated onto a grid in the Matlab software package. 

In addition to spatially mapping each strain component, knowledge of the displacement gradient matrix can be used to map the density of vortices and antivortices. Antivortices are distinguished from vortices by the chirality of the rotation in configuration space as you make a loop in real space. To quantify the chirality, we can compare the sign of the cross product of a pair of real-space vectors to their corresponding vectors in configuration space. Consider the real space cartesian vectors $\bf{x}$ and $\bf{y}$ where $\bf{x} \times \bf{y}$ is positive. They correspond to $\overline{d}\bf{x}$ and $\overline{d}\bf{y}$ in configuration space, by Eq. \ref{eq:d}.
 If the sign of the cross product in configuration space is also positive, it is a vortex. If negative, it is an antivortex. 

Given that $\bf{x}$ and $\bf{y}$ are basis vectors and the matrix \begin{equation*}
    \overline{d}=\left(	\begin{array}{cc}
	a & b \\
	c & d  \\	
	\end{array}
	\right) = \left(	\begin{array}{cc}
	\alpha+\beta & \gamma-\theta \\
	\gamma + \theta & \alpha-\beta  \\	
	\end{array}
	\right),
\end{equation*}      
$\overline{d}{\bf{x}} \times \overline{d}{\bf{y}}=(ad-cb) x y $. Thus the condition for a vortex is det$[\overline{d}]>0$ and for antivortex is det$[\overline{d}]<0$. 
	
	Writing in terms of the strain components, the condition is \begin{equation}
	    \text{sgn}[(\alpha^2 + \theta^2) - (\beta^2+\gamma^2)]= \begin{cases} 
       1 & \text{vortex} \\
       -1 & \text{antivortex}.
   \end{cases}
	\end{equation}

\bibliography{references.bib}

\providecommand{\noopsort}[1]{}\providecommand{\singleletter}[1]{#1}%
\begin{thebibliography}{33}%
\makeatletter
\providecommand \@ifxundefined [1]{%
 \@ifx{#1\undefined}
}%
\providecommand \@ifnum [1]{%
 \ifnum #1\expandafter \@firstoftwo
 \else \expandafter \@secondoftwo
 \fi
}%
\providecommand \@ifx [1]{%
 \ifx #1\expandafter \@firstoftwo
 \else \expandafter \@secondoftwo
 \fi
}%
\providecommand \natexlab [1]{#1}%
\providecommand \enquote  [1]{``#1''}%
\providecommand \bibnamefont  [1]{#1}%
\providecommand \bibfnamefont [1]{#1}%
\providecommand \citenamefont [1]{#1}%
\providecommand \href@noop [0]{\@secondoftwo}%
\providecommand \href [0]{\begingroup \@sanitize@url \@href}%
\providecommand \@href[1]{\@@startlink{#1}\@@href}%
\providecommand \@@href[1]{\endgroup#1\@@endlink}%
\providecommand \@sanitize@url [0]{\catcode `\\12\catcode `\$12\catcode
  `\&12\catcode `\#12\catcode `\^12\catcode `\_12\catcode `\%12\relax}%
\providecommand \@@startlink[1]{}%
\providecommand \@@endlink[0]{}%
\providecommand \url  [0]{\begingroup\@sanitize@url \@url }%
\providecommand \@url [1]{\endgroup\@href {#1}{\urlprefix }}%
\providecommand \urlprefix  [0]{URL }%
\providecommand \Eprint [0]{\href }%
\providecommand \doibase [0]{https://doi.org/}%
\providecommand \selectlanguage [0]{\@gobble}%
\providecommand \bibinfo  [0]{\@secondoftwo}%
\providecommand \bibfield  [0]{\@secondoftwo}%
\providecommand \translation [1]{[#1]}%
\providecommand \BibitemOpen [0]{}%
\providecommand \bibitemStop [0]{}%
\providecommand \bibitemNoStop [0]{.\EOS\space}%
\providecommand \EOS [0]{\spacefactor3000\relax}%
\providecommand \BibitemShut  [1]{\csname bibitem#1\endcsname}%
\let\auto@bib@innerbib\@empty
\bibitem [{\citenamefont {Ponomarenko}\ \emph {et~al.}(2013)\citenamefont
  {Ponomarenko}, \citenamefont {Gorbachev}, \citenamefont {Yu}, \citenamefont
  {Elias}, \citenamefont {Jalil}, \citenamefont {Patel}, \citenamefont
  {Mishchenko}, \citenamefont {Mayorov}, \citenamefont {Woods}, \citenamefont
  {Wallbank}, \citenamefont {Mucha-Kruczynski}, \citenamefont {Piot},
  \citenamefont {Potemski}, \citenamefont {Grigorieva}, \citenamefont
  {Novoselov}, \citenamefont {Guinea}, \citenamefont {Falko},\ and\
  \citenamefont {Geim}}]{HofPonomarenko}%
  \BibitemOpen
  \bibfield  {author} {\bibinfo {author} {\bibfnamefont {L.~A.}\ \bibnamefont
  {Ponomarenko}}, \bibinfo {author} {\bibfnamefont {R.~V.}\ \bibnamefont
  {Gorbachev}}, \bibinfo {author} {\bibfnamefont {G.~L.}\ \bibnamefont {Yu}},
  \bibinfo {author} {\bibfnamefont {D.~C.}\ \bibnamefont {Elias}}, \bibinfo
  {author} {\bibfnamefont {R.}~\bibnamefont {Jalil}}, \bibinfo {author}
  {\bibfnamefont {A.~A.}\ \bibnamefont {Patel}}, \bibinfo {author}
  {\bibfnamefont {A.}~\bibnamefont {Mishchenko}}, \bibinfo {author}
  {\bibfnamefont {A.~S.}\ \bibnamefont {Mayorov}}, \bibinfo {author}
  {\bibfnamefont {C.~R.}\ \bibnamefont {Woods}}, \bibinfo {author}
  {\bibfnamefont {J.~R.}\ \bibnamefont {Wallbank}}, \bibinfo {author}
  {\bibfnamefont {M.}~\bibnamefont {Mucha-Kruczynski}}, \bibinfo {author}
  {\bibfnamefont {B.~A.}\ \bibnamefont {Piot}}, \bibinfo {author}
  {\bibfnamefont {M.}~\bibnamefont {Potemski}}, \bibinfo {author}
  {\bibfnamefont {I.~V.}\ \bibnamefont {Grigorieva}}, \bibinfo {author}
  {\bibfnamefont {K.~S.}\ \bibnamefont {Novoselov}}, \bibinfo {author}
  {\bibfnamefont {F.}~\bibnamefont {Guinea}}, \bibinfo {author} {\bibfnamefont
  {V.~I.}\ \bibnamefont {Falko}},\ and\ \bibinfo {author} {\bibfnamefont
  {A.~K.}\ \bibnamefont {Geim}},\ }\bibfield  {title} {\bibinfo {title}
  {Cloning of dirac fermions in graphene superlattices},\ }\href@noop {}
  {\bibfield  {journal} {\bibinfo  {journal} {Nature}\ }\textbf {\bibinfo
  {volume} {497}},\ \bibinfo {pages} {594} (\bibinfo {year}
  {2013})}\BibitemShut {NoStop}%
\bibitem [{\citenamefont {Dean}\ \emph {et~al.}(2013)\citenamefont {Dean},
  \citenamefont {Wang}, \citenamefont {Maher}, \citenamefont {Forsythe},
  \citenamefont {Ghahari}, \citenamefont {Y.Gao}, \citenamefont {Katoch},
  \citenamefont {Ishigami}, \citenamefont {Moon}, \citenamefont {Koshino},
  \citenamefont {Taniguchi}, \citenamefont {Watanabe}, \citenamefont {Shepard},
  \citenamefont {J.Hone},\ and\ \citenamefont {Kim}}]{HofDean}%
  \BibitemOpen
  \bibfield  {author} {\bibinfo {author} {\bibfnamefont {C.~R.}\ \bibnamefont
  {Dean}}, \bibinfo {author} {\bibfnamefont {L.}~\bibnamefont {Wang}}, \bibinfo
  {author} {\bibfnamefont {P.}~\bibnamefont {Maher}}, \bibinfo {author}
  {\bibfnamefont {C.}~\bibnamefont {Forsythe}}, \bibinfo {author}
  {\bibfnamefont {F.}~\bibnamefont {Ghahari}}, \bibinfo {author} {\bibnamefont
  {Y.Gao}}, \bibinfo {author} {\bibfnamefont {J.}~\bibnamefont {Katoch}},
  \bibinfo {author} {\bibfnamefont {M.}~\bibnamefont {Ishigami}}, \bibinfo
  {author} {\bibfnamefont {P.}~\bibnamefont {Moon}}, \bibinfo {author}
  {\bibfnamefont {M.}~\bibnamefont {Koshino}}, \bibinfo {author} {\bibfnamefont
  {T.}~\bibnamefont {Taniguchi}}, \bibinfo {author} {\bibfnamefont
  {K.}~\bibnamefont {Watanabe}}, \bibinfo {author} {\bibfnamefont {K.~L.}\
  \bibnamefont {Shepard}}, \bibinfo {author} {\bibnamefont {J.Hone}},\ and\
  \bibinfo {author} {\bibfnamefont {P.}~\bibnamefont {Kim}},\ }\bibfield
  {title} {\bibinfo {title} {Hofstadter's butterfly and the fractal quantum
  hall effect in moire superlattices},\ }\href@noop {} {\bibfield  {journal}
  {\bibinfo  {journal} {Nature}\ }\textbf {\bibinfo {volume} {497}},\ \bibinfo
  {pages} {598} (\bibinfo {year} {2013})}\BibitemShut {NoStop}%
\bibitem [{\citenamefont {Hunt}\ \emph {et~al.}(2013)\citenamefont {Hunt},
  \citenamefont {Sanchez-Yamagishi}, \citenamefont {Young}, \citenamefont
  {Yankowitz}, \citenamefont {LeRoy}, \citenamefont {Watanabe}, \citenamefont
  {Taniguchi}, \citenamefont {Moon}, \citenamefont {Koshino}, \citenamefont
  {Jarillo-Herrero},\ and\ \citenamefont {Ashoori}}]{HofHunt}%
  \BibitemOpen
  \bibfield  {author} {\bibinfo {author} {\bibfnamefont {B.}~\bibnamefont
  {Hunt}}, \bibinfo {author} {\bibfnamefont {J.~D.}\ \bibnamefont
  {Sanchez-Yamagishi}}, \bibinfo {author} {\bibfnamefont {A.~F.}\ \bibnamefont
  {Young}}, \bibinfo {author} {\bibfnamefont {M.}~\bibnamefont {Yankowitz}},
  \bibinfo {author} {\bibfnamefont {B.~J.}\ \bibnamefont {LeRoy}}, \bibinfo
  {author} {\bibfnamefont {K.}~\bibnamefont {Watanabe}}, \bibinfo {author}
  {\bibfnamefont {T.}~\bibnamefont {Taniguchi}}, \bibinfo {author}
  {\bibfnamefont {P.}~\bibnamefont {Moon}}, \bibinfo {author} {\bibfnamefont
  {M.}~\bibnamefont {Koshino}}, \bibinfo {author} {\bibfnamefont
  {P.}~\bibnamefont {Jarillo-Herrero}},\ and\ \bibinfo {author} {\bibfnamefont
  {R.~C.}\ \bibnamefont {Ashoori}},\ }\bibfield  {title} {\bibinfo {title}
  {Massive dirac fermions and hofstadter butterfly in a van der waals
  heterostructure},\ }\href@noop {} {\bibfield  {journal} {\bibinfo  {journal}
  {Science}\ }\textbf {\bibinfo {volume} {340}},\ \bibinfo {pages} {1427}
  (\bibinfo {year} {2013})}\BibitemShut {NoStop}%
\bibitem [{\citenamefont {Gorbachev}\ \emph {et~al.}(2014)\citenamefont
  {Gorbachev}, \citenamefont {Song}, \citenamefont {Yu}, \citenamefont
  {Kretinin}, \citenamefont {Withers}, \citenamefont {Cao}, \citenamefont
  {Mishchenko}, \citenamefont {Grigorieva}, \citenamefont {Novoselov},
  \citenamefont {Levitov},\ and\ \citenamefont {Geim}}]{ValleyGorb}%
  \BibitemOpen
  \bibfield  {author} {\bibinfo {author} {\bibfnamefont {R.~V.}\ \bibnamefont
  {Gorbachev}}, \bibinfo {author} {\bibfnamefont {J.~C.~W.}\ \bibnamefont
  {Song}}, \bibinfo {author} {\bibfnamefont {G.~L.}\ \bibnamefont {Yu}},
  \bibinfo {author} {\bibfnamefont {A.~V.}\ \bibnamefont {Kretinin}}, \bibinfo
  {author} {\bibfnamefont {F.}~\bibnamefont {Withers}}, \bibinfo {author}
  {\bibfnamefont {Y.}~\bibnamefont {Cao}}, \bibinfo {author} {\bibfnamefont
  {A.}~\bibnamefont {Mishchenko}}, \bibinfo {author} {\bibfnamefont {I.~V.}\
  \bibnamefont {Grigorieva}}, \bibinfo {author} {\bibfnamefont {K.~S.}\
  \bibnamefont {Novoselov}}, \bibinfo {author} {\bibfnamefont {L.~S.}\
  \bibnamefont {Levitov}},\ and\ \bibinfo {author} {\bibfnamefont {A.~K.}\
  \bibnamefont {Geim}},\ }\bibfield  {title} {\bibinfo {title} {Detecting
  topological currents in graphene superlattices},\ }\href@noop {} {\bibfield
  {journal} {\bibinfo  {journal} {Science}\ }\textbf {\bibinfo {volume}
  {346}},\ \bibinfo {pages} {448} (\bibinfo {year} {2014})}\BibitemShut
  {NoStop}%
\bibitem [{\citenamefont {Endo}\ \emph {et~al.}(2019)\citenamefont {Endo},
  \citenamefont {Komatsu}, \citenamefont {Iwasaki}, \citenamefont {Watanabe},
  \citenamefont {Tsuya}, \citenamefont {Watanabe}, \citenamefont {Taniguchi},
  \citenamefont {Noguchi}, \citenamefont {Wakayama}, \citenamefont {Morita},\
  and\ \citenamefont {Moriyama}}]{ValleyEndo}%
  \BibitemOpen
  \bibfield  {author} {\bibinfo {author} {\bibfnamefont {K.}~\bibnamefont
  {Endo}}, \bibinfo {author} {\bibfnamefont {K.}~\bibnamefont {Komatsu}},
  \bibinfo {author} {\bibfnamefont {T.}~\bibnamefont {Iwasaki}}, \bibinfo
  {author} {\bibfnamefont {E.}~\bibnamefont {Watanabe}}, \bibinfo {author}
  {\bibfnamefont {D.}~\bibnamefont {Tsuya}}, \bibinfo {author} {\bibfnamefont
  {K.}~\bibnamefont {Watanabe}}, \bibinfo {author} {\bibfnamefont
  {T.}~\bibnamefont {Taniguchi}}, \bibinfo {author} {\bibfnamefont
  {Y.}~\bibnamefont {Noguchi}}, \bibinfo {author} {\bibfnamefont
  {Y.}~\bibnamefont {Wakayama}}, \bibinfo {author} {\bibfnamefont
  {Y.}~\bibnamefont {Morita}},\ and\ \bibinfo {author} {\bibfnamefont
  {S.}~\bibnamefont {Moriyama}},\ }\bibfield  {title} {\bibinfo {title}
  {Topological valley currents in bilayer graphene/hexagonal boron nitride
  superlattices},\ }\href@noop {} {\bibfield  {journal} {\bibinfo  {journal}
  {Appl. Phys. Lett.}\ }\textbf {\bibinfo {volume} {114}} (\bibinfo {year}
  {2019})}\BibitemShut {NoStop}%
\bibitem [{\citenamefont {Cao}\ \emph {et~al.}(2018{\natexlab{a}})\citenamefont
  {Cao}, \citenamefont {Fatemi}, \citenamefont {Demir}, \citenamefont {Fang},
  \citenamefont {Tomarken}, \citenamefont {Luo}, \citenamefont
  {Sanchez-Yamagishi}, \citenamefont {Watanabe}, \citenamefont {Taniguchi},
  \citenamefont {Kaxiras}, \citenamefont {Ashoori},\ and\ \citenamefont
  {Jarillo-Herrero}}]{CaoMott}%
  \BibitemOpen
  \bibfield  {author} {\bibinfo {author} {\bibfnamefont {Y.}~\bibnamefont
  {Cao}}, \bibinfo {author} {\bibfnamefont {V.}~\bibnamefont {Fatemi}},
  \bibinfo {author} {\bibfnamefont {A.}~\bibnamefont {Demir}}, \bibinfo
  {author} {\bibfnamefont {S.}~\bibnamefont {Fang}}, \bibinfo {author}
  {\bibfnamefont {S.~L.}\ \bibnamefont {Tomarken}}, \bibinfo {author}
  {\bibfnamefont {J.~Y.}\ \bibnamefont {Luo}}, \bibinfo {author} {\bibfnamefont
  {J.~D.}\ \bibnamefont {Sanchez-Yamagishi}}, \bibinfo {author} {\bibfnamefont
  {K.}~\bibnamefont {Watanabe}}, \bibinfo {author} {\bibfnamefont
  {T.}~\bibnamefont {Taniguchi}}, \bibinfo {author} {\bibfnamefont
  {E.}~\bibnamefont {Kaxiras}}, \bibinfo {author} {\bibfnamefont {R.~C.}\
  \bibnamefont {Ashoori}},\ and\ \bibinfo {author} {\bibfnamefont
  {P.}~\bibnamefont {Jarillo-Herrero}},\ }\bibfield  {title} {\bibinfo {title}
  {Unconventional superconductivity in magic-angle graphene superlattices},\
  }\href@noop {} {\bibfield  {journal} {\bibinfo  {journal} {Nature}\ }\textbf
  {\bibinfo {volume} {556}},\ \bibinfo {pages} {80} (\bibinfo {year}
  {2018}{\natexlab{a}})}\BibitemShut {NoStop}%
\bibitem [{\citenamefont {Cao}\ \emph {et~al.}(2018{\natexlab{b}})\citenamefont
  {Cao}, \citenamefont {Fatemi}, \citenamefont {Fang}, \citenamefont
  {Watanabe}, \citenamefont {Taniguchi}, \citenamefont {Kaxiras},\ and\
  \citenamefont {Jarillo-Herrero}}]{CaoMagic}%
  \BibitemOpen
  \bibfield  {author} {\bibinfo {author} {\bibfnamefont {Y.}~\bibnamefont
  {Cao}}, \bibinfo {author} {\bibfnamefont {V.}~\bibnamefont {Fatemi}},
  \bibinfo {author} {\bibfnamefont {S.}~\bibnamefont {Fang}}, \bibinfo {author}
  {\bibfnamefont {K.}~\bibnamefont {Watanabe}}, \bibinfo {author}
  {\bibfnamefont {T.}~\bibnamefont {Taniguchi}}, \bibinfo {author}
  {\bibfnamefont {E.}~\bibnamefont {Kaxiras}},\ and\ \bibinfo {author}
  {\bibfnamefont {P.}~\bibnamefont {Jarillo-Herrero}},\ }\bibfield  {title}
  {\bibinfo {title} {Correlated insulator behaviour at half-filling in
  magic-angle graphene superlattices},\ }\href@noop {} {\bibfield  {journal}
  {\bibinfo  {journal} {Nature}\ }\textbf {\bibinfo {volume} {556}},\ \bibinfo
  {pages} {43} (\bibinfo {year} {2018}{\natexlab{b}})}\BibitemShut {NoStop}%
\bibitem [{\citenamefont {Liu}\ \emph {et~al.}(2020)\citenamefont {Liu},
  \citenamefont {Hao}, \citenamefont {Khalaf}, \citenamefont {Lee},
  \citenamefont {Ronen}, \citenamefont {Yoo}, \citenamefont {Haei~Najafabadi},
  \citenamefont {Watanabe}, \citenamefont {Taniguchi}, \citenamefont
  {Vishwanath},\ and\ \citenamefont {Kim}}]{Xiaomeng-TDBG}%
  \BibitemOpen
  \bibfield  {author} {\bibinfo {author} {\bibfnamefont {X.}~\bibnamefont
  {Liu}}, \bibinfo {author} {\bibfnamefont {Z.}~\bibnamefont {Hao}}, \bibinfo
  {author} {\bibfnamefont {E.}~\bibnamefont {Khalaf}}, \bibinfo {author}
  {\bibfnamefont {J.~Y.}\ \bibnamefont {Lee}}, \bibinfo {author} {\bibfnamefont
  {Y.}~\bibnamefont {Ronen}}, \bibinfo {author} {\bibfnamefont
  {H.}~\bibnamefont {Yoo}}, \bibinfo {author} {\bibfnamefont {D.}~\bibnamefont
  {Haei~Najafabadi}}, \bibinfo {author} {\bibfnamefont {K.}~\bibnamefont
  {Watanabe}}, \bibinfo {author} {\bibfnamefont {T.}~\bibnamefont {Taniguchi}},
  \bibinfo {author} {\bibfnamefont {A.}~\bibnamefont {Vishwanath}},\ and\
  \bibinfo {author} {\bibfnamefont {P.}~\bibnamefont {Kim}},\ }\bibfield
  {title} {\bibinfo {title} {Tunable spin-polarized correlated states in
  twisted double bilayer graphene},\ }\href@noop {} {\bibfield  {journal}
  {\bibinfo  {journal} {Nature}\ }\textbf {\bibinfo {volume} {583}},\ \bibinfo
  {pages} {221} (\bibinfo {year} {2020})}\BibitemShut {NoStop}%
\bibitem [{\citenamefont {Zhang}\ \emph {et~al.}(2021)\citenamefont {Zhang},
  \citenamefont {Tsai}, \citenamefont {Zhu}, \citenamefont {Ren}, \citenamefont
  {Luo}, \citenamefont {Carr}, \citenamefont {Luskin}, \citenamefont
  {Kaxiras},\ and\ \citenamefont {Wang}}]{Ke}%
  \BibitemOpen
  \bibfield  {author} {\bibinfo {author} {\bibfnamefont {X.}~\bibnamefont
  {Zhang}}, \bibinfo {author} {\bibfnamefont {K.-T.}\ \bibnamefont {Tsai}},
  \bibinfo {author} {\bibfnamefont {Z.}~\bibnamefont {Zhu}}, \bibinfo {author}
  {\bibfnamefont {W.}~\bibnamefont {Ren}}, \bibinfo {author} {\bibfnamefont
  {Y.}~\bibnamefont {Luo}}, \bibinfo {author} {\bibfnamefont {S.}~\bibnamefont
  {Carr}}, \bibinfo {author} {\bibfnamefont {M.}~\bibnamefont {Luskin}},
  \bibinfo {author} {\bibfnamefont {E.}~\bibnamefont {Kaxiras}},\ and\ \bibinfo
  {author} {\bibfnamefont {K.}~\bibnamefont {Wang}},\ }\bibfield  {title}
  {\bibinfo {title} {Correlated insulating states and transport signature of
  superconductivity in twisted trilayer graphene superlattices},\ }\href@noop
  {} {\bibfield  {journal} {\bibinfo  {journal} {Physical Review Letters}\
  }\textbf {\bibinfo {volume} {127}} (\bibinfo {year} {2021})}\BibitemShut
  {NoStop}%
\bibitem [{\citenamefont {Hao}\ \emph {et~al.}(2021)\citenamefont {Hao},
  \citenamefont {Zimmerman}, \citenamefont {Ledwith}, \citenamefont {Khalaf},
  \citenamefont {Najafabadi}, \citenamefont {Watanabe}, \citenamefont
  {Taniguchi}, \citenamefont {Vishwanath},\ and\ \citenamefont
  {Kim}}]{ZeyuTrilayer}%
  \BibitemOpen
  \bibfield  {author} {\bibinfo {author} {\bibfnamefont {Z.}~\bibnamefont
  {Hao}}, \bibinfo {author} {\bibfnamefont {A.~M.}\ \bibnamefont {Zimmerman}},
  \bibinfo {author} {\bibfnamefont {P.}~\bibnamefont {Ledwith}}, \bibinfo
  {author} {\bibfnamefont {E.}~\bibnamefont {Khalaf}}, \bibinfo {author}
  {\bibfnamefont {D.~H.}\ \bibnamefont {Najafabadi}}, \bibinfo {author}
  {\bibfnamefont {K.}~\bibnamefont {Watanabe}}, \bibinfo {author}
  {\bibfnamefont {T.}~\bibnamefont {Taniguchi}}, \bibinfo {author}
  {\bibfnamefont {A.}~\bibnamefont {Vishwanath}},\ and\ \bibinfo {author}
  {\bibfnamefont {P.}~\bibnamefont {Kim}},\ }\bibfield  {title} {\bibinfo
  {title} {Electric field tunable superconductivity in alternating-twist
  magic-angle trilayer graphene},\ }\href@noop {} {\bibfield  {journal}
  {\bibinfo  {journal} {Science}\ }\textbf {\bibinfo {volume} {371}},\ \bibinfo
  {pages} {1133} (\bibinfo {year} {2021})}\BibitemShut {NoStop}%
\bibitem [{\citenamefont {Park}\ \emph {et~al.}(2021)\citenamefont {Park},
  \citenamefont {Cao}, \citenamefont {Watanabe}, \citenamefont {Taniguchi},\
  and\ \citenamefont {Jarillo-Herrero}}]{TrilayerPark}%
  \BibitemOpen
  \bibfield  {author} {\bibinfo {author} {\bibfnamefont {J.~M.}\ \bibnamefont
  {Park}}, \bibinfo {author} {\bibfnamefont {Y.}~\bibnamefont {Cao}}, \bibinfo
  {author} {\bibfnamefont {K.}~\bibnamefont {Watanabe}}, \bibinfo {author}
  {\bibfnamefont {T.}~\bibnamefont {Taniguchi}},\ and\ \bibinfo {author}
  {\bibfnamefont {P.}~\bibnamefont {Jarillo-Herrero}},\ }\bibfield  {title}
  {\bibinfo {title} {Tunable strongly coupled superconductivity in magic-angle
  twisted trilayer graphene},\ }\href@noop {} {\bibfield  {journal} {\bibinfo
  {journal} {Nature}\ }\textbf {\bibinfo {volume} {590}},\ \bibinfo {pages}
  {249} (\bibinfo {year} {2021})}\BibitemShut {NoStop}%
\bibitem [{\citenamefont {Park}\ \emph {et~al.}(2022)\citenamefont {Park},
  \citenamefont {Cao}, \citenamefont {Xia}, \citenamefont {Sun}, \citenamefont
  {Watanabe}, \citenamefont {Taniguchi},\ and\ \citenamefont
  {Jarillo-Herrero}}]{Quadrilayer}%
  \BibitemOpen
  \bibfield  {author} {\bibinfo {author} {\bibfnamefont {J.~M.}\ \bibnamefont
  {Park}}, \bibinfo {author} {\bibfnamefont {Y.}~\bibnamefont {Cao}}, \bibinfo
  {author} {\bibfnamefont {L.-Q.}\ \bibnamefont {Xia}}, \bibinfo {author}
  {\bibfnamefont {S.}~\bibnamefont {Sun}}, \bibinfo {author} {\bibfnamefont
  {K.}~\bibnamefont {Watanabe}}, \bibinfo {author} {\bibfnamefont
  {T.}~\bibnamefont {Taniguchi}},\ and\ \bibinfo {author} {\bibfnamefont
  {P.}~\bibnamefont {Jarillo-Herrero}},\ }\bibfield  {title} {\bibinfo {title}
  {Robust superconductivity in magic-angle multilayer graphene family},\
  }\href@noop {} {\bibfield  {journal} {\bibinfo  {journal} {Nature Materials}\
  ,\ \bibinfo {pages} {https://doi.org/10.1038/s41563}} (\bibinfo {year}
  {2022})}\BibitemShut {NoStop}%
\bibitem [{\citenamefont {Woods}\ \emph {et~al.}(2021)\citenamefont {Woods},
  \citenamefont {Ares}, \citenamefont {Nevison-Andrews}, \citenamefont
  {Holwill}, \citenamefont {Fabregas}, \citenamefont {Guinea}, \citenamefont
  {Geim}, \citenamefont {Novoselov}, \citenamefont {Walet},\ and\ \citenamefont
  {Fumagalli}}]{FE2}%
  \BibitemOpen
  \bibfield  {author} {\bibinfo {author} {\bibfnamefont {C.~R.}\ \bibnamefont
  {Woods}}, \bibinfo {author} {\bibfnamefont {P.}~\bibnamefont {Ares}},
  \bibinfo {author} {\bibfnamefont {H.}~\bibnamefont {Nevison-Andrews}},
  \bibinfo {author} {\bibfnamefont {M.~J.}\ \bibnamefont {Holwill}}, \bibinfo
  {author} {\bibfnamefont {R.}~\bibnamefont {Fabregas}}, \bibinfo {author}
  {\bibfnamefont {F.}~\bibnamefont {Guinea}}, \bibinfo {author} {\bibfnamefont
  {A.~K.}\ \bibnamefont {Geim}}, \bibinfo {author} {\bibfnamefont {K.~S.}\
  \bibnamefont {Novoselov}}, \bibinfo {author} {\bibfnamefont {N.~R.}\
  \bibnamefont {Walet}},\ and\ \bibinfo {author} {\bibfnamefont
  {L.}~\bibnamefont {Fumagalli}},\ }\bibfield  {title} {\bibinfo {title}
  {Charge-polarized interfacial superlattices in marginally twisted hexagonal
  boron nitride},\ }\href@noop {} {\bibfield  {journal} {\bibinfo  {journal}
  {Nature Communications}\ }\textbf {\bibinfo {volume} {12}} (\bibinfo {year}
  {2021})}\BibitemShut {NoStop}%
\bibitem [{\citenamefont {Wang}\ \emph {et~al.}(2022)\citenamefont {Wang},
  \citenamefont {Yasuda}, \citenamefont {Zhang}, \citenamefont {Liu},
  \citenamefont {Watanabe}, \citenamefont {Taniguchi}, \citenamefont
  {ad~Liang~Fu},\ and\ \citenamefont {Jarillo-Herrero}}]{FE3}%
  \BibitemOpen
  \bibfield  {author} {\bibinfo {author} {\bibfnamefont {X.}~\bibnamefont
  {Wang}}, \bibinfo {author} {\bibfnamefont {K.}~\bibnamefont {Yasuda}},
  \bibinfo {author} {\bibfnamefont {Y.}~\bibnamefont {Zhang}}, \bibinfo
  {author} {\bibfnamefont {S.}~\bibnamefont {Liu}}, \bibinfo {author}
  {\bibfnamefont {K.}~\bibnamefont {Watanabe}}, \bibinfo {author}
  {\bibfnamefont {T.}~\bibnamefont {Taniguchi}}, \bibinfo {author}
  {\bibfnamefont {J.~H.}\ \bibnamefont {ad~Liang~Fu}},\ and\ \bibinfo {author}
  {\bibfnamefont {P.}~\bibnamefont {Jarillo-Herrero}},\ }\bibfield  {title}
  {\bibinfo {title} {Interfacial ferroelectricity in rhombohedral stacked
  bilayer transition metal dichalcogenides},\ }\href@noop {} {\bibfield
  {journal} {\bibinfo  {journal} {Nature Nanotechnology}\ }\textbf {\bibinfo
  {volume} {17}},\ \bibinfo {pages} {367} (\bibinfo {year} {2022})}\BibitemShut
  {NoStop}%
\bibitem [{\citenamefont {Lau}\ \emph {et~al.}(2022)\citenamefont {Lau},
  \citenamefont {Bockrath}, \citenamefont {Mak},\ and\ \citenamefont
  {Zhang}}]{StrainReview}%
  \BibitemOpen
  \bibfield  {author} {\bibinfo {author} {\bibfnamefont {C.~N.}\ \bibnamefont
  {Lau}}, \bibinfo {author} {\bibfnamefont {M.~W.}\ \bibnamefont {Bockrath}},
  \bibinfo {author} {\bibfnamefont {K.~F.}\ \bibnamefont {Mak}},\ and\ \bibinfo
  {author} {\bibfnamefont {F.}~\bibnamefont {Zhang}},\ }\bibfield  {title}
  {\bibinfo {title} {Reproducibility in the fabrication and physics of moir\'e
  materials},\ }\href@noop {} {\bibfield  {journal} {\bibinfo  {journal}
  {Nature}\ }\textbf {\bibinfo {volume} {602}},\ \bibinfo {pages} {41}
  (\bibinfo {year} {2022})}\BibitemShut {NoStop}%
\bibitem [{\citenamefont {Uri}\ \emph {et~al.}(2020)\citenamefont {Uri},
  \citenamefont {Grover}, \citenamefont {Cao}, \citenamefont {Crosse},
  \citenamefont {Bagani}, \citenamefont {Rodan-Legrain}, \citenamefont
  {Myasoedov}, \citenamefont {Watanabe}, \citenamefont {Taniguchi},
  \citenamefont {Moon}, \citenamefont {Koshino}, \citenamefont
  {Jarillo-Herrero},\ and\ \citenamefont {Zeldov}}]{StrainmapSQUID}%
  \BibitemOpen
  \bibfield  {author} {\bibinfo {author} {\bibfnamefont {A.}~\bibnamefont
  {Uri}}, \bibinfo {author} {\bibfnamefont {S.}~\bibnamefont {Grover}},
  \bibinfo {author} {\bibfnamefont {Y.}~\bibnamefont {Cao}}, \bibinfo {author}
  {\bibfnamefont {J.~A.}\ \bibnamefont {Crosse}}, \bibinfo {author}
  {\bibfnamefont {K.}~\bibnamefont {Bagani}}, \bibinfo {author} {\bibfnamefont
  {D.}~\bibnamefont {Rodan-Legrain}}, \bibinfo {author} {\bibfnamefont
  {Y.}~\bibnamefont {Myasoedov}}, \bibinfo {author} {\bibfnamefont
  {K.}~\bibnamefont {Watanabe}}, \bibinfo {author} {\bibfnamefont
  {T.}~\bibnamefont {Taniguchi}}, \bibinfo {author} {\bibfnamefont
  {P.}~\bibnamefont {Moon}}, \bibinfo {author} {\bibfnamefont {M.}~\bibnamefont
  {Koshino}}, \bibinfo {author} {\bibfnamefont {P.}~\bibnamefont
  {Jarillo-Herrero}},\ and\ \bibinfo {author} {\bibfnamefont {E.}~\bibnamefont
  {Zeldov}},\ }\bibfield  {title} {\bibinfo {title} {Mapping the twist-angle
  disorder and landau levels in magic-angle graphene},\ }\href@noop {}
  {\bibfield  {journal} {\bibinfo  {journal} {Nature}\ }\textbf {\bibinfo
  {volume} {581}},\ \bibinfo {pages} {47} (\bibinfo {year} {2020})}\BibitemShut
  {NoStop}%
\bibitem [{\citenamefont {Wilson}\ \emph {et~al.}(2020)\citenamefont {Wilson},
  \citenamefont {Fu}, \citenamefont {{Das Sarma}},\ and\ \citenamefont
  {Pixley}}]{Strain2}%
  \BibitemOpen
  \bibfield  {author} {\bibinfo {author} {\bibfnamefont {J.~H.}\ \bibnamefont
  {Wilson}}, \bibinfo {author} {\bibfnamefont {Y.}~\bibnamefont {Fu}}, \bibinfo
  {author} {\bibfnamefont {S.}~\bibnamefont {{Das Sarma}}},\ and\ \bibinfo
  {author} {\bibfnamefont {J.~H.}\ \bibnamefont {Pixley}},\ }\bibfield  {title}
  {\bibinfo {title} {Disorder in twisted bilayer graphene},\ }\href@noop {}
  {\bibfield  {journal} {\bibinfo  {journal} {Phys. Rev. Research}\ }\textbf
  {\bibinfo {volume} {2}} (\bibinfo {year} {2020})}\BibitemShut {NoStop}%
\bibitem [{\citenamefont {Bai}\ \emph {et~al.}(2020)\citenamefont {Bai},
  \citenamefont {Zhou}, \citenamefont {Wang}, \citenamefont {Wu}, \citenamefont
  {McGilly}, \citenamefont {Halbertal}, \citenamefont {Lo}, \citenamefont
  {Liu}, \citenamefont {Ardelean}, \citenamefont {Rivera}, \citenamefont
  {Finney}, \citenamefont {Yang}, \citenamefont {Basov}, \citenamefont {Yao},
  \citenamefont {Xu}, \citenamefont {Hone}, \citenamefont {Pasupathy},\ and\
  \citenamefont {Zhu}}]{Excitons}%
  \BibitemOpen
  \bibfield  {author} {\bibinfo {author} {\bibfnamefont {Y.}~\bibnamefont
  {Bai}}, \bibinfo {author} {\bibfnamefont {L.}~\bibnamefont {Zhou}}, \bibinfo
  {author} {\bibfnamefont {J.}~\bibnamefont {Wang}}, \bibinfo {author}
  {\bibfnamefont {W.}~\bibnamefont {Wu}}, \bibinfo {author} {\bibfnamefont
  {L.~J.}\ \bibnamefont {McGilly}}, \bibinfo {author} {\bibfnamefont
  {D.}~\bibnamefont {Halbertal}}, \bibinfo {author} {\bibfnamefont {C.~F.~B.}\
  \bibnamefont {Lo}}, \bibinfo {author} {\bibfnamefont {F.}~\bibnamefont
  {Liu}}, \bibinfo {author} {\bibfnamefont {J.}~\bibnamefont {Ardelean}},
  \bibinfo {author} {\bibfnamefont {P.}~\bibnamefont {Rivera}}, \bibinfo
  {author} {\bibfnamefont {N.~R.}\ \bibnamefont {Finney}}, \bibinfo {author}
  {\bibfnamefont {X.-C.}\ \bibnamefont {Yang}}, \bibinfo {author}
  {\bibfnamefont {D.~N.}\ \bibnamefont {Basov}}, \bibinfo {author}
  {\bibfnamefont {W.}~\bibnamefont {Yao}}, \bibinfo {author} {\bibfnamefont
  {X.}~\bibnamefont {Xu}}, \bibinfo {author} {\bibfnamefont {J.}~\bibnamefont
  {Hone}}, \bibinfo {author} {\bibfnamefont {A.~N.}\ \bibnamefont
  {Pasupathy}},\ and\ \bibinfo {author} {\bibfnamefont {X.-Y.}\ \bibnamefont
  {Zhu}},\ }\bibfield  {title} {\bibinfo {title} {Excitons in strain-induced
  one-dimensional moir\'e potentials at transition metal dichalcogenide
  heterojunctions},\ }\href@noop {} {\bibfield  {journal} {\bibinfo  {journal}
  {Nat. Mater.}\ }\textbf {\bibinfo {volume} {10}},\ \bibinfo {pages} {1124}
  (\bibinfo {year} {2020})}\BibitemShut {NoStop}%
\bibitem [{\citenamefont {Yoo}\ \emph {et~al.}(2019)\citenamefont {Yoo},
  \citenamefont {Engelke}, \citenamefont {Carr}, \citenamefont {Fang},
  \citenamefont {Zhang}, \citenamefont {Cazeaux}, \citenamefont {Sung},
  \citenamefont {Hovden}, \citenamefont {Tsen}, \citenamefont {Taniguchi},
  \citenamefont {Watanabe}, \citenamefont {Yi}, \citenamefont {Kim},
  \citenamefont {Luskin}, \citenamefont {Tadmor}, \citenamefont {Kaxiras},\
  and\ \citenamefont {Kim}}]{Yoo}%
  \BibitemOpen
  \bibfield  {author} {\bibinfo {author} {\bibfnamefont {H.}~\bibnamefont
  {Yoo}}, \bibinfo {author} {\bibfnamefont {R.}~\bibnamefont {Engelke}},
  \bibinfo {author} {\bibfnamefont {S.}~\bibnamefont {Carr}}, \bibinfo {author}
  {\bibfnamefont {S.}~\bibnamefont {Fang}}, \bibinfo {author} {\bibfnamefont
  {K.}~\bibnamefont {Zhang}}, \bibinfo {author} {\bibfnamefont
  {P.}~\bibnamefont {Cazeaux}}, \bibinfo {author} {\bibfnamefont {S.~H.}\
  \bibnamefont {Sung}}, \bibinfo {author} {\bibfnamefont {R.}~\bibnamefont
  {Hovden}}, \bibinfo {author} {\bibfnamefont {A.~W.}\ \bibnamefont {Tsen}},
  \bibinfo {author} {\bibfnamefont {T.}~\bibnamefont {Taniguchi}}, \bibinfo
  {author} {\bibfnamefont {K.}~\bibnamefont {Watanabe}}, \bibinfo {author}
  {\bibfnamefont {G.-C.}\ \bibnamefont {Yi}}, \bibinfo {author} {\bibfnamefont
  {M.}~\bibnamefont {Kim}}, \bibinfo {author} {\bibfnamefont {M.}~\bibnamefont
  {Luskin}}, \bibinfo {author} {\bibfnamefont {E.~B.}\ \bibnamefont {Tadmor}},
  \bibinfo {author} {\bibfnamefont {E.}~\bibnamefont {Kaxiras}},\ and\ \bibinfo
  {author} {\bibfnamefont {P.}~\bibnamefont {Kim}},\ }\bibfield  {title}
  {\bibinfo {title} {Atomic and electronic reconstruction at the van der waals
  interface in twisted bilayer graphene},\ }\href@noop {} {\bibfield  {journal}
  {\bibinfo  {journal} {Nature Materials}\ }\textbf {\bibinfo {volume} {18}},\
  \bibinfo {pages} {448} (\bibinfo {year} {2019})}\BibitemShut {NoStop}%
\bibitem [{\citenamefont {Alden}\ \emph {et~al.}(2013)\citenamefont {Alden},
  \citenamefont {Tsen}, \citenamefont {Huang}, \citenamefont {Hovden},
  \citenamefont {Brown}, \citenamefont {Park}, \citenamefont {Muller},\ and\
  \citenamefont {McEuen}}]{Alden}%
  \BibitemOpen
  \bibfield  {author} {\bibinfo {author} {\bibfnamefont {J.~S.}\ \bibnamefont
  {Alden}}, \bibinfo {author} {\bibfnamefont {A.~W.}\ \bibnamefont {Tsen}},
  \bibinfo {author} {\bibfnamefont {P.~Y.}\ \bibnamefont {Huang}}, \bibinfo
  {author} {\bibfnamefont {R.}~\bibnamefont {Hovden}}, \bibinfo {author}
  {\bibfnamefont {L.}~\bibnamefont {Brown}}, \bibinfo {author} {\bibfnamefont
  {J.}~\bibnamefont {Park}}, \bibinfo {author} {\bibfnamefont {D.~A.}\
  \bibnamefont {Muller}},\ and\ \bibinfo {author} {\bibfnamefont {P.~L.}\
  \bibnamefont {McEuen}},\ }\bibfield  {title} {\bibinfo {title} {Strain
  solitons and topological defects in bilayer graphene},\ }\href@noop {}
  {\bibfield  {journal} {\bibinfo  {journal} {Proceedings of the National
  Academy of Sciences of the USA}\ }\textbf {\bibinfo {volume} {110}},\
  \bibinfo {pages} {11256} (\bibinfo {year} {2013})}\BibitemShut {NoStop}%
\bibitem [{\citenamefont {Turkel}\ \emph {et~al.}(2022)\citenamefont {Turkel},
  \citenamefont {Swann}, \citenamefont {Zhu}, \citenamefont {Christos},
  \citenamefont {Watanabe}, \citenamefont {Taniguchi}, \citenamefont {Sachdev},
  \citenamefont {Scheurer}, \citenamefont {Kaxiras}, \citenamefont {Dean},\
  and\ \citenamefont {Pasupathy}}]{Turkel}%
  \BibitemOpen
  \bibfield  {author} {\bibinfo {author} {\bibfnamefont {S.}~\bibnamefont
  {Turkel}}, \bibinfo {author} {\bibfnamefont {J.}~\bibnamefont {Swann}},
  \bibinfo {author} {\bibfnamefont {Z.}~\bibnamefont {Zhu}}, \bibinfo {author}
  {\bibfnamefont {M.}~\bibnamefont {Christos}}, \bibinfo {author}
  {\bibfnamefont {K.}~\bibnamefont {Watanabe}}, \bibinfo {author}
  {\bibfnamefont {T.}~\bibnamefont {Taniguchi}}, \bibinfo {author}
  {\bibfnamefont {S.}~\bibnamefont {Sachdev}}, \bibinfo {author} {\bibfnamefont
  {M.~S.}\ \bibnamefont {Scheurer}}, \bibinfo {author} {\bibfnamefont
  {E.}~\bibnamefont {Kaxiras}}, \bibinfo {author} {\bibfnamefont {C.~R.}\
  \bibnamefont {Dean}},\ and\ \bibinfo {author} {\bibfnamefont {A.~N.}\
  \bibnamefont {Pasupathy}},\ }\bibfield  {title} {\bibinfo {title} {Orderly
  disorder in magic-angle twisted trilayer graphene},\ }\href@noop {}
  {\bibfield  {journal} {\bibinfo  {journal} {Science}\ }\textbf {\bibinfo
  {volume} {376}},\ \bibinfo {pages} {193} (\bibinfo {year}
  {2022})}\BibitemShut {NoStop}%
\bibitem [{\citenamefont {Yu}\ \emph {et~al.}(2020)\citenamefont {Yu},
  \citenamefont {Zhang}, \citenamefont {Parks}, \citenamefont {Babar},
  \citenamefont {andIsaac M.~Craig}, \citenamefont {Winkle}, \citenamefont
  {Lyssenko}, \citenamefont {Taniguchi}, \citenamefont {Watanabe},
  \citenamefont {Viswanathan},\ and\ \citenamefont
  {Bediako}}]{Kwabena-chemistry}%
  \BibitemOpen
  \bibfield  {author} {\bibinfo {author} {\bibfnamefont {Y.}~\bibnamefont
  {Yu}}, \bibinfo {author} {\bibfnamefont {K.}~\bibnamefont {Zhang}}, \bibinfo
  {author} {\bibfnamefont {H.}~\bibnamefont {Parks}}, \bibinfo {author}
  {\bibfnamefont {M.}~\bibnamefont {Babar}}, \bibinfo {author} {\bibfnamefont
  {S.~C.}\ \bibnamefont {andIsaac M.~Craig}}, \bibinfo {author} {\bibfnamefont
  {M.~V.}\ \bibnamefont {Winkle}}, \bibinfo {author} {\bibfnamefont
  {A.}~\bibnamefont {Lyssenko}}, \bibinfo {author} {\bibfnamefont
  {T.}~\bibnamefont {Taniguchi}}, \bibinfo {author} {\bibfnamefont
  {K.}~\bibnamefont {Watanabe}}, \bibinfo {author} {\bibfnamefont
  {V.}~\bibnamefont {Viswanathan}},\ and\ \bibinfo {author} {\bibfnamefont
  {D.~K.}\ \bibnamefont {Bediako}},\ }\bibfield  {title} {\bibinfo {title}
  {Tunable angle-dependent electrochemistry at twisted bilayer graphene with
  moir\'e flat bands},\ }\href@noop {} {\bibfield  {journal} {\bibinfo
  {journal} {Nature Chemistry}\ }\textbf {\bibinfo {volume} {14}},\ \bibinfo
  {pages} {267} (\bibinfo {year} {2020})}\BibitemShut {NoStop}%
\bibitem [{\citenamefont {Thouless}(1998)}]{thouless_book}%
  \BibitemOpen
  \bibfield  {author} {\bibinfo {author} {\bibfnamefont {D.}~\bibnamefont
  {Thouless}},\ }\href@noop {} {\emph {\bibinfo {title} {Topological quantum
  numbers in nonrelativistic physics}}}\ (\bibinfo  {publisher} {World
  Scientific},\ \bibinfo {year} {1998})\BibitemShut {NoStop}%
\bibitem [{\citenamefont {Mermin}(1979)}]{Mermin}%
  \BibitemOpen
  \bibfield  {author} {\bibinfo {author} {\bibfnamefont {N.~D.}\ \bibnamefont
  {Mermin}},\ }\bibfield  {title} {\bibinfo {title} {The topological theory of
  defects in ordered media},\ }\href@noop {} {\bibfield  {journal} {\bibinfo
  {journal} {Rev. Mod. Phys.}\ }\textbf {\bibinfo {volume} {51}},\ \bibinfo
  {pages} {591} (\bibinfo {year} {1979})}\BibitemShut {NoStop}%
\bibitem [{\citenamefont {Carr}\ \emph {et~al.}(2018)\citenamefont {Carr},
  \citenamefont {Massatt}, \citenamefont {Torrisi}, \citenamefont {Cazeaux},
  \citenamefont {Luskin},\ and\ \citenamefont {Kaxiras}}]{CarrRelaxation}%
  \BibitemOpen
  \bibfield  {author} {\bibinfo {author} {\bibfnamefont {S.}~\bibnamefont
  {Carr}}, \bibinfo {author} {\bibfnamefont {D.}~\bibnamefont {Massatt}},
  \bibinfo {author} {\bibfnamefont {S.~B.}\ \bibnamefont {Torrisi}}, \bibinfo
  {author} {\bibfnamefont {P.}~\bibnamefont {Cazeaux}}, \bibinfo {author}
  {\bibfnamefont {M.}~\bibnamefont {Luskin}},\ and\ \bibinfo {author}
  {\bibfnamefont {E.}~\bibnamefont {Kaxiras}},\ }\bibfield  {title} {\bibinfo
  {title} {Relaxation and domain formation in incommensurate two-dimensional
  heterostructures},\ }\href@noop {} {\bibfield  {journal} {\bibinfo  {journal}
  {Physical Review B}\ }\textbf {\bibinfo {volume} {98}} (\bibinfo {year}
  {2018})}\BibitemShut {NoStop}%
\bibitem [{\citenamefont {Lin}\ \emph {et~al.}(2013)\citenamefont {Lin},
  \citenamefont {Fang}, \citenamefont {Zhou}, \citenamefont {Lupin},
  \citenamefont {Idrobo}, \citenamefont {Kong}, \citenamefont {Pennycook}, ,\
  and\ \citenamefont {Pantelide}}]{LinAC}%
  \BibitemOpen
  \bibfield  {author} {\bibinfo {author} {\bibfnamefont {J.}~\bibnamefont
  {Lin}}, \bibinfo {author} {\bibfnamefont {W.}~\bibnamefont {Fang}}, \bibinfo
  {author} {\bibfnamefont {W.}~\bibnamefont {Zhou}}, \bibinfo {author}
  {\bibfnamefont {A.~R.}\ \bibnamefont {Lupin}}, \bibinfo {author}
  {\bibfnamefont {J.~C.}\ \bibnamefont {Idrobo}}, \bibinfo {author}
  {\bibfnamefont {J.}~\bibnamefont {Kong}}, \bibinfo {author} {\bibfnamefont
  {S.~J.}\ \bibnamefont {Pennycook}}, ,\ and\ \bibinfo {author} {\bibfnamefont
  {S.~T.}\ \bibnamefont {Pantelide}},\ }\bibfield  {title} {\bibinfo {title}
  {{{AC}}/{{AB}} stacking boundaries in bilayer graphene},\ }\href@noop {}
  {\bibfield  {journal} {\bibinfo  {journal} {Nano Letters}\ }\textbf {\bibinfo
  {volume} {13}},\ \bibinfo {pages} {3262} (\bibinfo {year}
  {2013})}\BibitemShut {NoStop}%
\bibitem [{\citenamefont {Friedrichs}(1947)}]{StrainMatrix}%
  \BibitemOpen
  \bibfield  {author} {\bibinfo {author} {\bibfnamefont {K.~O.}\ \bibnamefont
  {Friedrichs}},\ }\bibfield  {title} {\bibinfo {title} {On the boundary-value
  problems of the theory of elasticity and korn's inequality},\ }\href@noop {}
  {\bibfield  {journal} {\bibinfo  {journal} {Annals of Mathematics}\ }\textbf
  {\bibinfo {volume} {48}},\ \bibinfo {pages} {441} (\bibinfo {year}
  {1947})}\BibitemShut {NoStop}%
\bibitem [{\citenamefont {Cohen}(1989)}]{free-group}%
  \BibitemOpen
  \bibfield  {author} {\bibinfo {author} {\bibfnamefont {D.~E.}\ \bibnamefont
  {Cohen}},\ }\href@noop {} {\emph {\bibinfo {title} {Combinatorial Group
  Theory: A Topological Approach (London Mathematical Society Student Texts,
  Series Number 14)}}}\ (\bibinfo  {publisher} {Cambridge University Press},\
  \bibinfo {year} {1989})\BibitemShut {NoStop}%
\bibitem [{\citenamefont {Kharlampovich}\ and\ \citenamefont
  {Myasnikov}(2006)}]{free-nonabelian-group}%
  \BibitemOpen
  \bibfield  {author} {\bibinfo {author} {\bibfnamefont {O.}~\bibnamefont
  {Kharlampovich}}\ and\ \bibinfo {author} {\bibfnamefont {A.}~\bibnamefont
  {Myasnikov}},\ }\bibfield  {title} {\bibinfo {title} {Elementary theory of
  free non-abelian groups},\ }\href
  {https://doi.org/https://doi.org/10.1016/j.jalgebra.2006.03.033} {\bibfield
  {journal} {\bibinfo  {journal} {Journal of Algebra}\ }\textbf {\bibinfo
  {volume} {302}},\ \bibinfo {pages} {451} (\bibinfo {year}
  {2006})}\BibitemShut {NoStop}%
\bibitem [{\citenamefont {Guinea}\ \emph {et~al.}(2010)\citenamefont {Guinea},
  \citenamefont {Katsnelson},\ and\ \citenamefont {Geim}}]{PseudoGeim}%
  \BibitemOpen
  \bibfield  {author} {\bibinfo {author} {\bibfnamefont {F.}~\bibnamefont
  {Guinea}}, \bibinfo {author} {\bibfnamefont {M.~I.}\ \bibnamefont
  {Katsnelson}},\ and\ \bibinfo {author} {\bibfnamefont {A.~K.}\ \bibnamefont
  {Geim}},\ }\bibfield  {title} {\bibinfo {title} {Energy gaps and a zero-field
  quantum hall effect in graphene by strain engineering},\ }\href@noop {}
  {\bibfield  {journal} {\bibinfo  {journal} {Nature Physics}\ }\textbf
  {\bibinfo {volume} {6}},\ \bibinfo {pages} {30} (\bibinfo {year}
  {2010})}\BibitemShut {NoStop}%
\bibitem [{\citenamefont {Halbertal}\ \emph {et~al.}(2022)\citenamefont
  {Halbertal}, \citenamefont {Shabani}, \citenamefont {Passupathy}, ,\ and\
  \citenamefont {Basov}}]{DorriStrain}%
  \BibitemOpen
  \bibfield  {author} {\bibinfo {author} {\bibfnamefont {D.}~\bibnamefont
  {Halbertal}}, \bibinfo {author} {\bibfnamefont {S.}~\bibnamefont {Shabani}},
  \bibinfo {author} {\bibfnamefont {A.~N.}\ \bibnamefont {Passupathy}}, ,\ and\
  \bibinfo {author} {\bibfnamefont {D.~N.}\ \bibnamefont {Basov}},\ }\bibfield
  {title} {\bibinfo {title} {Moir\'e metrology of energy landscapes in van der
  waals heterostructures},\ }\href@noop {} {\bibfield  {journal} {\bibinfo
  {journal} {ACS Nano}\ }\textbf {\bibinfo {volume} {16}},\ \bibinfo {pages}
  {1471} (\bibinfo {year} {2022})}\BibitemShut {NoStop}%
\bibitem [{\citenamefont {Badia}\ and\ \citenamefont
  {Verdugo}(2020)}]{Badia2020}%
  \BibitemOpen
  \bibfield  {author} {\bibinfo {author} {\bibfnamefont {S.}~\bibnamefont
  {Badia}}\ and\ \bibinfo {author} {\bibfnamefont {F.}~\bibnamefont
  {Verdugo}},\ }\bibfield  {title} {\bibinfo {title} {Gridap: An extensible
  finite element toolbox in julia},\ }\href@noop {} {\bibfield  {journal}
  {\bibinfo  {journal} {Journal of Open Source Software}\ }\textbf {\bibinfo
  {volume} {5}},\ \bibinfo {pages} {2520} (\bibinfo {year} {2020})}\BibitemShut
  {NoStop}%
\bibitem [{\citenamefont {Geuzaine}\ and\ \citenamefont
  {Remacle}(2009)}]{Gmsh}%
  \BibitemOpen
  \bibfield  {author} {\bibinfo {author} {\bibfnamefont {C.}~\bibnamefont
  {Geuzaine}}\ and\ \bibinfo {author} {\bibfnamefont {J.-F.}\ \bibnamefont
  {Remacle}},\ }\bibfield  {title} {\bibinfo {title} {Gmsh: A 3-{{D}} finite
  element mesh generator with built-in pre- and post-processing facilities:
  {{THE GMSH PAPER}}},\ }\href@noop {} {\bibfield  {journal} {\bibinfo
  {journal} {International Journal on Numerical Methods in Engineering}\
  }\textbf {\bibinfo {volume} {79}},\ \bibinfo {pages} {1309} (\bibinfo {year}
  {2009})}\BibitemShut {NoStop}%
\end{thebibliography}%

\end{document}